\documentclass[useAMS,usenatbib]{mn2e}
\usepackage{txfonts}
\usepackage{amssymb}
\usepackage[dvips]{graphicx}
\usepackage{lscape}
\usepackage{color}
\title[Predicting the fate of binary red giants]
{Predicting the Fate of Binary Red Giants Using the Observed Sequence E Star Population:
Binary Planetary Nebula Nuclei and Post-RGB Stars}

\author[J. D. Nie et al.]
{J. D. Nie$^{1,2}$
\thanks{E-mail:niejundan@mail.bnu.edu.cn(JDN);wood@mso.anu.edu.au(PRW); cnicholls@physics.ucsd.edu (CPN)},
P. R. Wood$^{2}$\footnotemark[1] and C. P. Nicholls$^{3}$\footnotemark[1]\\
$^{1}$Department of Astronomy, Beijing Normal University, Beijing, 100875, China\\
$^{2}$Research School of Astronomy and Astrophysics, Australian National
University, Cotter Road, Weston Creek ACT 2611, Australia\\
$^{3}$Center for Astrophysics and Space Science, University of California San Diego, 
La Jolla, CA 92093, USA}

\begin{document}

\date{Accepted; Received; in original form}

\pagerange{\pageref{firstpage}--\pageref{lastpage}} \pubyear{2011}

\maketitle

\label{firstpage}

\begin{abstract}\label{abstract}

Sequence E variables are close binary red giants that show ellipsoidal
light variations. They are likely the immediate precursors of
planetary nebulae (PNe) with close binary central, stars as well as
other binary post-AGB and binary post-RGB stars. We have made a Monte
Carlo simulation to determine the fraction of red giant binaries that
go through a common envelope (CE) event leading to the production of
a close binary system or a merged star. The novel aspect of this
simulation is that we use the observed frequency of sequence E
binaries in the LMC to normalize our calculations. This normalization
allows us to produce predictions that are relatively independent of
model assumptions. In our standard model, and assuming that the
relative numbers of PNe of various types are proportional to their
birthrates, we find that in the LMC today the fraction of PNe
with close binary central stars is 7--9\%, the fraction of PNe with
intermediate period binary central stars having separations
capable of influencing the nebula shape (orbital periods less than
500 years) is 23--27\%, the fraction of PNe containing wide
binaries that are unable to influence the nebula shape (orbital
period greater than 500 years) is 46--55\%, the fraction of PNe 
derived from  single stars is 3--19\% and 5-6\% of PNe are
produced by previously merged stars.  We also predict that the
birthrate of post-RGB stars is $\sim$4\% of the total PN
  birthrate, equivalent to $\sim$50\% of the production rate of PNe
with close binary central stars.  These post-RGB stars most likely
appear initially as luminous low-mass helium white dwarf binaries. The
average lifetime of sequence E ellipsoidal variability with amplitude
more than 0.02 magnitudes is predicted to be $\sim$0.95~Myr.  We use
our model and the observed number of red giant stars in the top one
magnitude of the RGB in the LMC to predict the number of PNe in 
the LMC.  We predict 548 PNe in good agreement with the
541$\pm$89 PNe observed by \citet{2006MNRAS.373..521R}.  Since most of
these PNe come from single or non-interacting binary stars in our
model, this means that most such stars produce PNe contrary to the
``Binary Hypothesis'' which suggests that binary interaction is
required to produce a PN.

\end{abstract}


\section{Introduction}\label{introduction}
In the LMC and SMC, red giant variables follow at least five Period-Luminosity 
relations, originally named sequences A to E \citep{1999IAUS..191..151W}. 
Sequence A, B and C are radial pulsation sequences and sequence D, which is 
parallel to sequences A, B and C but of longer period, is still of unknown 
physical origin \citep{2004ApJ...604..800W,2009MNRAS.399.2063N,2010AJ....139.1909N}. 
Sequence E stars follow a loose Period-Luminosity relation between sequences C 
and D, but they extend to lower luminosities. Sequence E stars are widely 
accepted to be binaries \citep{1999IAUS..191..151W,2004AcA....54..347S}.  
The orbital periods of the sequence E red giants are typically 100--600
days and the full light curve amplitudes are less than 0.6 mag in the 
MACHO red ($M_R$) band. They lie on the first giant branch in the case 
of low-mass stars, or the equivalent pre-core helium burning giant phase 
of intermediate mass stars \citep{1999IAUS..191..151W}. Some of the sequence 
E stars also extend to the AGB \citep{2004AcA....54..347S,2007ApJ...660.1486S}. 
As a whole, sequence E stars make up approximately 0.5--2\% of the RGB
and AGB stars in the Large Magellanic Cloud (LMC).

Sequence E variables show ellipsoidal light variations
\citep{2004AcA....54..347S}. In these systems, the primary star has
evolved to the red giant phase and substantially filled its Roche lobe. 
Because the orbital separation of the system is similar to the size of 
the red giant, the red giant is distorted from spherical symmetry to an 
ellipsoid or pear-like shape by tidal interactions with its unseen companion. 
It is the orbital rotation of this shape that gives rise to the ellipsoidal 
light variations.  About 7\% of sequence E variables show eclipses in 
addition to the ellipsoidal light variations \citep{2004AcA....54..347S}.

The first radial velocity curves of sequence E stars were presented in
\citet*{2010MNRAS.405.1770N} and from their observations, the mean full
velocity amplitude is about 43 km~s$^{-1}$.  This is consistent with
the amplitudes expected for close red giant binaries with roughly
solar-mass components. The velocity variation of the ellipsoidal variable 
is regular, dominated by its orbital motion.  For every single cycle of 
the velocity variation there are two cycles of light variation, the latter 
being caused by the change in apparent surface area.

In past and ongoing radial velocity studies of sequence E stars
in the LMC, we have obtained spectra for 110 systems.  In none of these
systems do we see the emission lines characteristic of symbiotic
stars.  This means that the companion is unlikely to be a white dwarf
or neutron star.  It is presumably a main-sequence star so that the 
red giant is usually the primary star in the binary system.

As the Roche lobe of the sequence E red variables is already substantially 
filled, further expansion of the red giant as it evolves is expected 
to soon cause Roche lobe overflow. The binary system will suffer a common 
envelope (CE) event (except in the rare case that the red giant has 
previously lost sufficient mass that mass transfer is stable). A CE event 
will lead to the ejection of the red giant envelope and the termination of 
RGB or AGB evolution. If the two stellar cores do not merge in the CE event, 
a luminous compact close binary system will be left. In the case of a CE event 
occurring on the RGB, the compact system will be a low mass white dwarf 
binary; while on the AGB, the remnant stellar core will be the central star 
of a planetary nebula (PN). If the two stellar cores do merge they should 
form a rapidly rotating red giant of the FK Comae class 
\citep{1976ApJ...209..829W,1981ApJ...247L.131B, 1981IAUS...93..177B}.
The fraction of the rapidly rotating red giants is about 2\% \citep
{2011ApJ...732...39C}, although many of these stars may be tidally
locked binaries rather than single stars.

The most common morphology of PNe is nonspherical and there are several 
theories attempting to explain the underlying mechanisms responsible for 
the shaping. Among all the mechanisms, the binary hypothesis is most favoured 
\citep*{1990ApJ...355..568B,1993ApJ...418..794Y,1997ApJS..112..487S, 
2000ASPC..199..115B, 2007BaltA..16...79Z,2009PASP..121..316D} and it has 
been confirmed by observations in some cases. In fact, the shape of PNe 
with close binary central stars is definitely observed to be asymmetric.
It has been suggested that binary interaction plays a major role in the
formation and shaping of many \citep{1990ApJ...355..568B,1993PASP..105.1373I}, 
or even most PNe \citep{2000ASPC..199..115B,2009PASP..121..316D}. 

At the present time, the actual fraction of PNe with close binary central 
stars is not clear.  The short periods of the known central stars of 
binary PNe mean that these stars have been through a CE phase since the 
companion currently lies well within the radius of the precursor star when 
it was a red giant. It seems fairly clear that the brighter sequence E 
stars are indeed the immediate precursors of the known binary PNe central stars. 
At lower luminosities, the low mass white dwarf
binaries, which are produced from RGB binaries \citep[e.g.][]{1976IAUS...73...75P,
1984ApJ...277..355W}, should be descendants of sequence E stars. 
It is our aim to use the observed fraction of sequence E stars among all red 
giants to estimate the fraction of PNe with close binary central stars.
We also use the sequence E fraction and the overall population of RGB stars in the
LMC to estimate the birthrate and population of all PNe there, along with the
birthrate of low mass white dwarf binaries.

In a study using MACHO 
light curves of stars in a central part of the LMC bar, \citet{1999IAUS..191..151W} 
found that $\sim$0.5\% of the red giants in the top one magnitude of the 
RGB were detected as sequence E variables in the $M_R$ band. Similarly,
\citet{2004AcA....54..347S} and \citet{2007ApJ...660.1486S} using OGLE 
II data found that 1--2\% of red giants were detected as sequence E 
variables. These numbers provide the calibrating input for the fractions of close 
binary PNe and low mass helium white dwarfs produced by CE events. Simultaneously, 
with this calibration, we can also estimate the fraction of red giants that 
are in wider binaries which reach the AGB tip without Roche lobe filling, 
and the fraction of PNe that are descended from single stars. 

\section[]{The Simulation Model}\label{simulation}

The prediction of the evolutionary fate of binary and single red
giants is made by a Monte Carlo simulation. One million red giant
binary systems are generated using orbital element distributions
derived from observations of stars in the solar vicinity. The
evolutionary fate of these binary systems is then examined.  Single
stars are added to the total stellar population in order to reproduce
the observed fraction of sequence E stars on the red giant branch.
Details of the standard simulation model are described in Section
\ref{standard}.  Besides the standard model, we adjust the inputs to
see how dependent the results are on the model assumptions.  Results
obtained by varying one or multiple parameters are described in
Section \ref{non-standard}. Since we will be utilizing the observed
sequence E population in the LMC, our model uses inputs from LMC
sources when available.

\subsection[]{The standard model}\label{standard}
\subsubsection[]{Generating the binary system}\label{generation}
\begin{enumerate}
\renewcommand{\theenumi}{(\arabic{enumi})}
\item The initial mass of the primary star \label{mass}\\
Initially we require the red giant to be the primary star in the binary systems. The 
primary mass is drawn from the initial mass distribution $h(m,t)$ defined by
\begin{equation}
dN=h(m,t)dmdt~~,
\end{equation}
where $h(m,t)$ is the number of stars born per unit mass per unit time. To 
obtain $h(m,t)$, we consider the Initial Mass Function (IMF) as well as the 
star formation history. Let $g(m)$ be the IMF (assumed independent of time) 
and let $f(t)$ be the total star formation rate at time $t$, then $h(m,t)$ 
is expressed as:
\begin{equation}
h(m,t)=g(m)\times f(t)~~.
\end{equation}
Here, $g(m)$ is assumed to follow the \cite{1955ApJ...121..161S} power law: 
$g(m)=m^{-2.35}$. An observationally-based approximation to the LMC star 
formation history \citep{1992ApJ...388..400B} has a star burst which starts 
$\sim$4 Gyrs ago and ceases $\sim$0.5 Gyrs ago, with the ratio of burst to 
quiescent star formation rate $R_{\rm burst}$  being 10. Thus, $f(t)$ is 
expressed as:
\[
f(t)= \left\{ \begin{array}{cc}
0~, & t<0.5\\
R_{\rm burst}\times{b}~, & 0.5<t<4~~,\\
b~, & t>4
\end{array} \right.
\]
where $R_{\rm burst}$=10, $b$ is a constant, and $t$ is in Gyrs from the present 
time. According to the evolutionary tracks of \citet{2000A&AS..141..371G}  
appropriate for the LMC (Z=0.008), 
after converting the ages of red giants to initial masses, we have:
\begin{equation}\label{rburst}
f(t(m))=k(m)= \left\{ \begin{array}{cc}
0~,& m>3.0\\
R_{\rm burst}\times{b}~,& 3.0>m>1.3~~,\\
b~,& m<1.3
\end{array} \right.
\end{equation}
where $m$ is in solar masses.

With the initial mass distribution, the cumulative probability distribution 
of the mass, $H(m)$, is expressed by:
\begin{equation}
\label{hm}
H(m)=\int_{m_0}^{m}{g(m)\times k(m)dm}~~,
\end{equation}
where $m_0$ corresponds to the oldest stars currently on the giant branch.

As we only consider red giant stars, the mass range should be that of RGB 
and AGB stars.  The evolutionary tracks of \citet{2000A&AS..141..371G} show 
that the minimum mass of low-mass stars in the LMC that are 
currently on the first giant branch is about $\rm{0.9~M_{\odot}}$. Stars 
with $m <\rm{0.9~M_{\odot}}$ have not yet evolved to the RGB since such 
stars have a main sequence lifetime longer than the time since the first 
stars formed. Hence, in Equation (\ref{hm}) we set $m_0$ to $\rm{0.9~M_{\odot}}$. 
The upper limiting mass for RGB stars which develop electron degenerate 
helium cores on the first ascent of giant branch, is about $\rm{1.85~M_{\odot}}$. 
Stars with mass larger than this do not reach the high luminosities of the 
observed sequence E stars in the LMC until they enter the early-AGB (EAGB). 
As the LMC star formation is assumed to cease 0.5 Gyrs ago, the maximum 
mass for AGB stars is $\rm{3.0~M_{\odot}}$. In the Monte Carlo simulation, 
we adjust `$b$' so that $H(3.0) = 1$, and draw the primary mass from the 
normalized $H(m)$.

\item The initial mass ratio  \label{mass_ratio}\\
We define $q=m_2/m_1$ as the mass ratio, where $m_1$ is the mass of the 
primary star and $m_2$ is the mass of the secondary star. 

The initial mass ratio is drawn from the distribution given in 
\citet{1991A&A...248..485D} who give orbital element distributions
for binary stars in the solar vicinity.
The mass ratio distribution follows a Gaussian-type relation, with a peak at 
0.23. The reason we use the distribution from the solar vicinity (as well as 
the orbital period distribution described below) is that 
the mass ratio distribution is unknown observationally in the LMC.

Given that we now know both masses in the binary system, we can decide
which component is `currently' being seen as the red giant.
This is done by randomly assigning the primary or the secondary as the
current red giant with probabilities that are proportional to the
star formation rate at the time the system would have been born
if the primary or secondary, respectively, was the current red giant.
In the case where the secondary is the current red giant, the 
full evolution of the primary is followed before the evolution of the
secondary is considered.

\item The orbital period \label{period}\\
The initial orbital period ($P$) is drawn from the distribution given in 
\citet{1991A&A...248..485D}. The period distribution follows a Gaussian-type 
relation, with a peak of log$P$=4.8, in unit of days. 

Note that some binaries with long periods are too wide  
to influence the shape of PNe.  \citet*{1998ApJ...496..842S} suggest 
$P<$ 500--2000 years if a wide binary is to influence a PN shape. 
\citet{1999ApJ...523..357M} similarly give a gravitational focusing 
fraction $\alpha_{foc}$ greater than 0.03 for a binary to produce a shell 
more distorted than ``quasi-spherical''. If we apply this to low-mass stars 
with masses of $\sim$$1.5~\rm{M_{\odot}}$ and wind velocities of 
$\sim$10 km s$^{-1}$, we find $P <$ several hundred years is required. 
Based on the above, we adopt 500 years as the upper limiting period 
 $P_{\rm max}$ for a wide binary to influence the shape of a PN.
Hereinafter, we refer to binaries with $P > P_{\rm max}$ as `wide
binaries' and binaries with $P < P_{\rm max}$ but with periods large
enough to avoid Roche lobe overflow on the RGB and AGB as `intermediate
period' binaries.

\item Eccentricity and stellar separation  \label{separation}\\
About 90\% of the observed light curves of sequence E stars indicate zero 
or small eccentricity \citep{2004AcA....54..347S}, so we assume zero 
eccentricity. Given the primary 
mass, mass ratio and orbital period, and assumed zero eccentricity, we can 
calculate the separation of the stellar centres $a$ via the equation
of orbital motion.   

\item  Orbital inclination \label{inclination}\\
The orbital inclination $i$ is obtained assuming a random orientation 
of the orbital pole so that $dN=\sin i\,di$.
\end{enumerate}

\subsubsection{Luminosity limits for RGB and AGB stars}\label{term_lim}
According to observational studies \citep*{1983ApJ...275..773F,
1999IAUS..191..151W, 2003MNRAS.343L..79K} and the the evolutionary 
tracks of \citet{2000A&AS..141..371G} for a typical LMC metallicity of Z=0.008, 
the bolometric magnitude for the RGB 
tip (TRGB) is close to $-3.6$ mag. Thus, we set 
${\emph M}_{\rm{bol}}{\rm (TRGB)} = -3.6$ as the maximum luminosity of 
sequence E stars on the RGB. In our simulation we compare the observed and 
simulated ratio of sequence E stars to all red giants at luminosities 
corresponding to the brightest one magnitude of the RGB, 
i.e. $-2.6 > M_{\rm bol} > -3.6$. We consider the evolution of stars once 
they become brighter than the base of the RGB.
 
AGB stars extend in luminosity up to the point at which a superwind rapidly 
terminates the AGB evolution.  The luminosity at the tip of the AGB, 
$M_{\rm bol}\rm{(TAGB)}$, is calculated from:
\begin{equation}
\label{lrgbt}
M_{\rm{bol}}\rm{(TAGB)}=-5.1-0.6(\emph{m}_{\rm 1,0}-1.55)~~.
\end{equation}
Here, the zero point $-5.1$ is obtained from the maximum luminosity for AGB 
stars in the rich intermediate-age LMC cluster NGC 1978 \citep{2010MNRAS.408..522K}.  
In this cluster, the mass of the O-rich stars on the RGB is $\sim$1.55~$\rm{M_{\odot}}$ 
\citep{2010MNRAS.408..522K}. In Equation (\ref{lrgbt}), $m_{\rm 1,0}$ is the initial mass
of the red giant.  The slope 0.6 is the AGB tip magnitude change per 
solar mass for stars in the 
range of 1.0--3.0~$\rm{M_{\odot}}$ \citep*{1990ApJ...352...96F,1994ApJS...92..125V}.  
Binary stars that evolve to the AGB tip without filling their Roche lobes are 
assumed to eject their envelopes rapidly by a superwind, in the same way as a 
single star.

\subsubsection[]{Mass loss from the red giants}\label{mass_loss}
We consider stellar mass loss for all evolving stars.  We use the
empirical formulation by \citet{1975MSRSL...8..369R} to calculate the
mass loss rate, but with the rate multiplied by a parameter $\eta$
which is set equal to 0.33 \citep{1983ARA&A..21..271I,2005A&A...441.1117L}.
 
According to \citet*{1988MNRAS.231..823T}, the mass loss rate may 
be tidally enhanced by a factor of:
\[1+B\times min [(\frac{R}{R_L})^6,\frac{1}{2^6}]~~,
\]
where $R$ is the radius of the primary star, 
$R_L$ is the equivalent radius of the Roche lobe. 
\citet*{1988MNRAS.231..823T} suggest $B$=10000. However, it is unlikely 
that such enhanced mass loss rates are realistic for a wide range of 
binary systems (see the references in Section~\ref{non-standard}), 
so we treat `$B$' as a variable parameter. 
In our standard model it is set to zero.

In order to
calculate the amount of mass lost per magnitude of evolution up the
giant branch, and to calculate the lifetimes of the sequence E stars,
we need the evolution rate d$M_{\rm{bol}}$/d$t$.  For low-mass stars 
with $m < 1.85$\,M$_{\odot}$ on the RGB, 
the evolutionary rate in the tracks of
\citet{2008A&A...484..815B} is well approximated (in mag/Myr) by
\begin{equation}
\frac{dM_{\rm bol}}{dt} = -7.58\times10^{-3}-2.96\times10^{-4}\frac{L}{{\rm L}_{\odot}}+4.73\times10^{-8}\left(\frac{L}{{\rm L}_{\odot}}\right)^2~.
\end{equation}
On the EAGB,
where the helium burning shell alone supplies most of the star's luminosity,
the evolution rate for all initial masses considered here (0.9--3.0 \,M$_{\odot}$) is 
approximated by
\begin{equation}
\frac{dM_{\rm bol}}{dt} = 1.15 + 0.75 M_{\rm bol}~~~. 
\end{equation}
On the AGB ($M_{\rm bol} < -3.6$), the average evolution rate of the
full AGB calculations of \citet{1994ApJS...92..125V} for LMC
metallicity is approximated for all masses in the range 
0.9--3.0 \,M$_{\odot}$ by
\begin{equation}
\frac{dM_{\rm bol}}{dt} = 1/(0.456-0.085M_{\rm bol}^2)~~~.
\end{equation}
 Although these formulae are derived from evolutionary models for
single stars, they also apply for red giants in binary systems
since the nuclear evolution of the red giant core is independent
of the conditions in the convective envelope 
\citep{1970A&A.....6..426R,1986ApJ...307..659W}.

Many binary stars in the simulation reach luminosities on the AGB where 
a superwind causes AGB termination, just as happens for a single star. 
This termination is assumed to occur over a brief interval and it is 
treated by specifying an AGB termination luminosity (see Section~\ref{term_lim}).

Since the primary star loses mass in a wind, the secondary star
can accrete some of the material as it orbits through it.  We
include wind accretion by the \citet{1944MNRAS.104..273B} mechanism, as
given by equation (6) of \citet*{2002MNRAS.329..897H} with $\alpha_W
= 3/2$ and $\beta_W = 1/2$.  The orbital evolution resulting from
mass loss and accretion is calculated using equation (20) in
\citet{2002MNRAS.329..897H}.

\subsubsection{Tidal effects}
When a red giant substantially fills its Roche lobe, tides induced in
the red giant by the orbiting companion cause the conversion of
orbital angular momentum into spin angular momentum of the red giant.
We treat this process using the equations in Section 2.3.1 of 
\citet{2002MNRAS.329..897H}.  In particular, the rate of change of
spin angular momentum is given by equations (26) and (35) of \citet{2002MNRAS.329..897H}, 
with the spin angular momentum change being extracted from orbital angular 
momentum of the binary system.  We also include the loss of
spin angular momentum from the red giant by stellar wind mass loss
using equation (11) of \citet{2002MNRAS.329..897H}.

\subsubsection[]{Light variation of an ellipsoidal variable}\label{variation}
For an ellipsoidal binary system containing a red giant, the amplitude 
of the light variation depends on the fractional filling of the Roche 
lobe by the red giant, the mass ratio and the orbital inclination.

\begin{figure}
\begin{center}
\includegraphics[angle=0,width=0.45\textwidth,height=0.7\hsize]{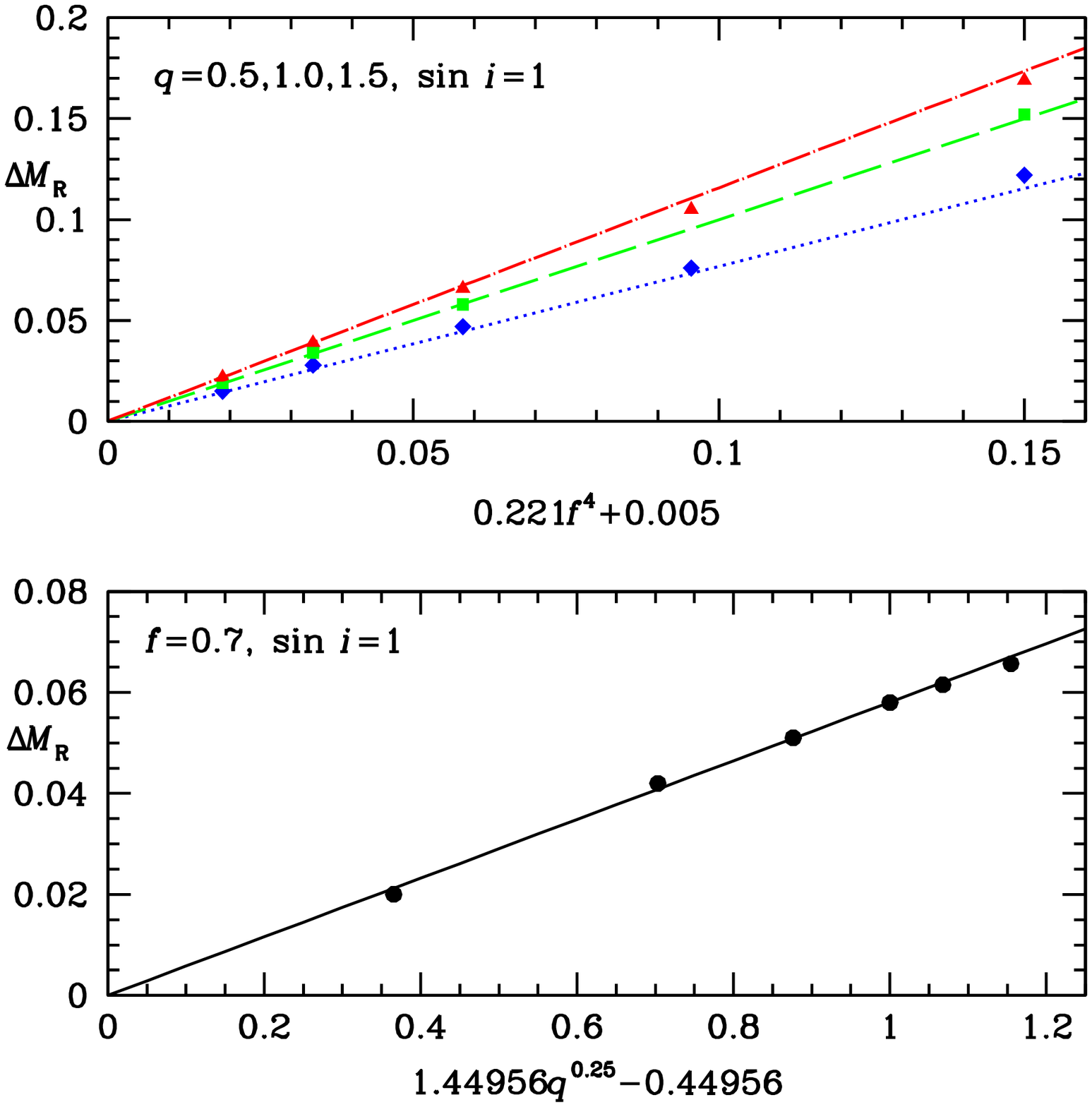}
\includegraphics[angle=0,width=0.45\textwidth,height=0.7\hsize]{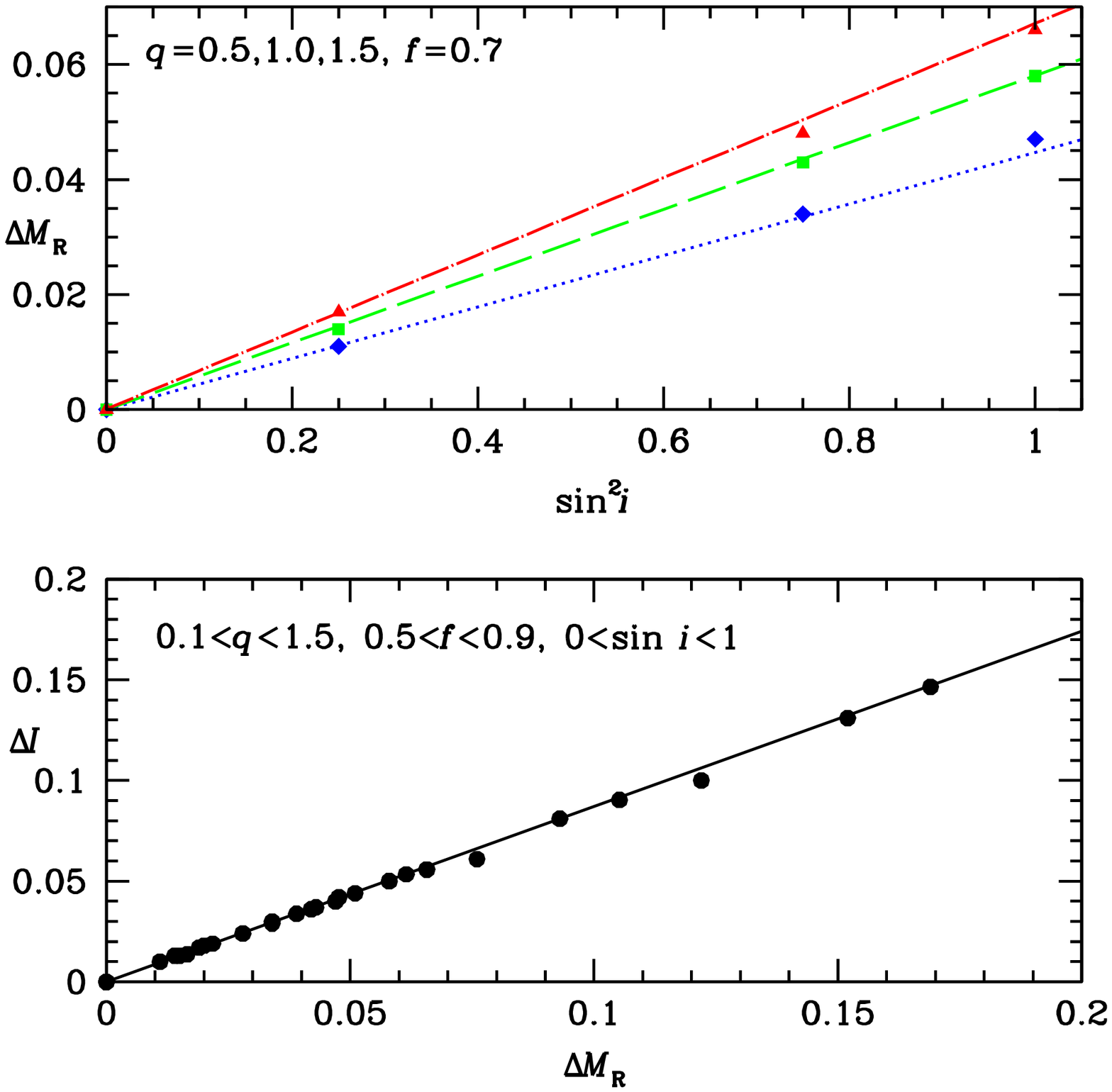}
\caption{Fits of $\Delta M_{R}$ to the factors in Equation (\ref{ramp}) 
(top three panels) and the relation between $\Delta I$ and $\Delta
M_{R}$ (bottom panel). In the first and third panels, blue diamonds denote 
the values of $\Delta M_{R}$ when $q=0.5$, green squares denote the values 
of $\Delta M_{R}$ when $q=1.0$, and the red triangles denote the values of 
$\Delta M_{R}$ when $q=1.5$. Line fits according to Equation (\ref{ramp}) 
are also shown, blue dotted lines correspond to blue diamonds, green
long dashed lines correspond to green squares, and the red 
dot-long dashed lines correspond to red triangles. In other 
panels, solid circles denote the values obtained from the \textsc{nightfall} 
program and the corresponding black lines are the fits. \label{fig_nightfall}}
\end{center}
\end{figure}

Rather than using an analytic approximation to the ellipsoidal light
variations \citep[e.g.][]{1985ApJ...295..143M}, we use the program
\textsc{nightfall}\footnote{http://www.hs.uni-hamburg.de/DE/Ins/Per/Wichmann/Nightfall.html}
to model the observed ellipsoidal light variations of partial-Roche
lobe filling systems as a function of binary parameters. In these
models, the effective temperature of the red giant is set to 4000 K,
as most of our sequence E stars are K or M type stars. A blackbody
flux distribution is assumed. Since the companions of the
  sequence E stars are expected to be mostly main sequence stars, the
radius and luminosity of the secondary star are set to small values so
they do not affect the light variation .  As noted in
  Section~\ref{generation}, we assume zero eccentricity. We then use
\textsc{nightfall} to create light curves for ellipsoidal variables
with a range of values for the Roche lobe radius filling factor $f$,
mass ratio $q$ and inclination $i$. Then full light amplitudes in the
$M_R$ and $I$ bands are derived, and a fit to the input parameters is
made. We find the full light variation amplitude in $M_R$ band is well
approximated by:
\begin{equation}
\label{ramp}
\Delta{M_R}=(0.221f^{4}+0.005)\times(1.44956q^{0.25}-0.44956)\times{sin^2i}~~,
\end{equation}
for $0.5<f<0.9,~ 0.1<q<1.5,~0 < i < \frac{\pi}{2}$. We also find the 
relation between  $\Delta{M_R}$ and $\Delta{I}$ is well approximated by:
\begin{equation}
\label{iamp}
\Delta{I}=0.87\times{\Delta{M_R}}~~,
\end{equation}
where $\Delta{I}$ is the full light curve amplitude in the $I$ band.
The lines in Fig. \ref{fig_nightfall} show fits to the factors in 
equations (\ref{ramp}) and (\ref{iamp}).

\subsubsection[]{Radii during the ellipsoidal phase of the red giant evolution}\label{radii}
Given a minimum amplitude for detectable light variations (see Section 
\ref{macho_ogle}), the minimum radius filling factor $f_{\rm min}$ of a 
red giant in a binary system with detectable ellipsoidal light variation 
can be determined from Equations (\ref{ramp}) and (\ref{iamp}) after $q$ 
and $i$ have been selected. The corresponding minimum red giant radius 
$R_{\rm min}$ is given by
\begin{equation}
\label{rmin}
R_{\rm{min}}=R_{\rm{L}}\times{f_{\rm min}}~~,
\end{equation}
where $R_{\rm{L}}$ is the equivalent radius of the Roche lobe \citep{1983ApJ...268..368E}:
\[
\frac{R_{\rm{L}}}{a}=\frac{0.49q^{-2/3}}{0.6q^{-2/3}+ln(1+q^{-1/3})}~.
\]
The maximum stellar radius $R_{\rm{max}}$, corresponding to the Roche lobe
being filled, will be
\begin{equation}
\label{rmax}
R_{\rm{max}}=R_{\rm{L}}~~.
\end{equation}

\subsubsection[]{Roche lobe overflow}\label{out_CE} 
The further expansion of a Roche lobe-filling red giant as it evolves
will cause Roche lobe overflow. Then, provided the mass ratio
$q=m_2/m_1$ is smaller than the critical mass ratio $q_{\rm crit}=(m_2/m_1)_{\rm crit}$
for the unstable mass transfer (see below), the binary components
will come into contact and undergo a CE event.

In binaries that undergo 
a CE event, there are two possible outcomes -- a close binary system or 
a single coalesced star.  Following widely used procedures 
\citep[e.g. see Section 2.7.1 of][]{2002MNRAS.329..897H}, 
we assume that the orbital energy dissipated in the interaction process 
goes into overcoming the binding energy of the red giant envelope which 
is thus lost.  If neither the remnant red giant core nor the companion 
main sequence star fills its Roche lobe at the end of this process, a 
close binary containing a white dwarf (the red giant core) and a main 
sequence star will be left. However, if the main sequence star fills its 
Roche lobe, then Roche lobe overflow will cause the main sequence star 
to mix with the common envelope. The red giant core thus gains H-rich 
material and the result will be a single, merged red giant whose further 
evolution will produce a PN with a single-star nucleus. Note that as the 
red giant core is much more compact than the main sequence star, it is 
unlikely that this core will fill its Roche lobe.  

The above possibilities are treated in our model using equations
(69)--(77) of \citet{2002MNRAS.329..897H}, with  $\lambda = 0.5$
and $\alpha_{\rm CE} = 1$ in our standard model.  According to
\citet{1990ApJ...358..189D,1992A&A...261..188D,1993ApJ...418..794Y}
and \citet*{1995MNRAS.272..800H}, $\alpha_{\rm CE}=1$ gives the best
fit to observations (we also try lower values - see Section~\ref
{non-standard}). We assume the mass of the red giant core is given
by the formulae $L/{\rm L_{\odot}} = 59250\times(M_{\rm c}/{\rm
  M_{\odot}}-0.48)$ for AGB stars and $L/{\rm L_{\odot}} =
5.3\times10^{5}(M_{\rm c}/{\rm M_{\odot}})^{6.667}$ for RGB stars, which are
the fits to the core mass--luminosity relation on the red giant branch
shown by the evolutionary tracks of \citet{2008A&A...484..815B} at
metallicity Z=0.008.

To check when stable mass transfer rather than a CE event occurs,
we use equation (57) of \citet{2002MNRAS.329..897H} to calculate
$q_{\rm crit}$.  Note that this equation assumes conservative mass
transfer which we also assume.  Except for small envelope masses (less
than 20\% of the red giant mass), the critical mass ratio
$(m_2/m_1)_{\rm crit}$ given by the above equation is greater than 1.0
and very few of red giants end up with stable mass transfer in our
standard model.  Very high mass loss rates or different values of
$q_{\rm crit}$ can give larger numbers of red giants with stable mass
transfer (see Section~\ref{non-standard} where we run models with a
lower value for $q_{\rm crit}$).

In binaries that undergo stable mass transfer, the red giant evolves
closely filling its Roche lobe and transferring matter to its
companion. Roche lobe filling is maintained in the models by enforcing
a large mass transfer rate when the red giant radius exceed its Roche
lobe radius.  This is done using equation (2) of
\citet{2008MNRAS.387.1416C}.

\subsubsection[]{Effective temperature and bolometric magnitude}\label{teff_lum}
Low-mass stars evolve through the RGB and later ascend the EAGB and AGB. When 
stars enter the EAGB and AGB they are oxygen rich. However, after several 
thermal pulses, which lead to dredge--up of carbon to the stellar atmosphere, 
the O-rich stars are converted to C-rich stars.  This leads to a decrease 
in effective temperature $T_{\rm eff}$.  Owing to this, $T_{\rm eff}$ values 
are calculated separately for these two sub-types.

\begin{figure}
\begin{center}
\includegraphics[width=84mm]{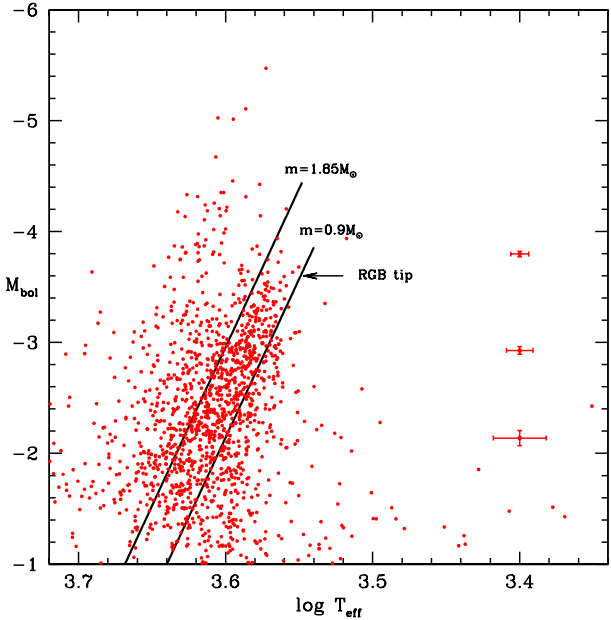}
\caption{The HR diagram of the sequence E stars. The red points are the 
observation data, while the black lines are the red giant branch fits for 
masses of $0.9~\rm{M_{\odot}}$ and $1.85~\rm{M_{\odot}}$. Typical 
observational error bars are shown at three magnitude levels. The maximum 
luminosity of each black line corresponds to the transition luminosity 
from O-rich to C-rich stars. The luminosity of the RGB tip is also marked.
\label{fig_hrd}}
\end{center}
\end{figure}

We use the intermediate age LMC globular cluster NGC 1978 as a template for 
LMC red giant properties. According to the data of \citet{2010MNRAS.408..522K}, 
the giant branch slope in the HR diagram for O-rich stars is given by 
$d \log T_{\rm eff}$/$dM_{\rm bol}=0.035$. Moreover, from the evolutionary 
tracks of \citet{2000A&AS..141..371G}, we find $d \log T_{\rm eff}$/$dm=0.03$ 
(with $m$ in solar masses). The effective temperature of O-rich stars is thus 
given by
\begin{equation}
\label{T-LM}
\log \emph{T}_{\rm eff}=\log \emph{T}_{\rm eff,0}+0.035M_{\rm bol}+0.03m_1~~. 
\end{equation}
To find the zero point $\log \emph{T}_{\rm eff,0}$ for the effective temperature, 
we fit Equation (\ref{T-LM}) to the observed HR diagram 
(Fig. \ref{fig_hrd}) for LMC sequence E stars in \citet{2004AcA....54..347S}. 
The effective temperatures of the observed sequence E stars are calculated by 
converting their $(J-K)_{0}$ to $T_{\rm{eff}}$ using the transforms in
\citet{2000AJ....119.1424H,2000AJ....119.1448H}. We adopt $E(B-V)$=0.08 
\citep{2006ApJ...642..834K}, $E(J-K)=0.52E(B-V)$ and $A(K)=0.35E(B-V)$ 
\citep{1985ApJ...288..618R}. The bolometric correction $BC_{K}$ is calculated 
from $(J-K)_{0}$ with the transforms in \citet{2000AJ....119.1424H,2000AJ....119.1448H}. 
The $J$ and $K$ magnitudes were obtained from the 2MASS catalogue 
\citep{2003yCat.2246....0C}. Fitting the RGB given by Equation (13) for 
masses between 0.9~$\rm{M_{\odot}}$ and 1.85~$\rm{M_{\odot}}$ 
to the densest part of the observed RGB in Fig. \ref{fig_hrd} gives $\log \emph{T}_{\rm eff,0}=3.648$.
Note that 0.9 and 1.85 $\rm{M_{\odot}}$ are the lower and upper mass 
limits, respectively, for low mass RGB stars (see Section~\ref{generation}).

There are substantial numbers of sequence E red giants in
Fig. \ref{fig_hrd} which are brighter than $M_{\rm bol} = -3$ and
which have $\log \emph{T}_{\rm eff}$ about 0.08 larger than that of
the low mass RGB stars.  These red giants are intermediate mass stars
($M > 1.85\,\rm{M_{\odot}}$) with non-degenerate He cores evolving to
the point of He core burning or perhaps on the EAGB. Other
sequence E red giants with large deviations in $\log \emph{T}_{\rm
  eff}$ from the main locus of the RGB in Fig. \ref{fig_hrd},
such as those in the lower right of the figure, 
presumably have photometry contaminated by spatially coincident field
stars of the LMC or the Galaxy or they may be high mass loss rate AGB
stars with thick obscuring circumstellar shells.

Because C-rich stars are often surrounded by substantial dust shells that cause 
an unknown amount of reddening, it is not possible to use observed photometric 
colors to derive the photospheric temperature. We therefore use red giant 
models to obtain the position of the C-rich star giant branch. For C-rich stars, 
the slope of the giant branch in the HR diagram is obtained from Table 5 in 
\citet{2010MNRAS.408..522K}. The models in this table were computed to fit 
the O-rich giant branch in NGC 1978, as well as allowing for C/O ratios to 
increase from unity at luminosities brighter than the transition luminosity 
for O to C-rich stars.  The models predict that 
$d \rm{log~\emph{T}_{eff}}$/$d \rm{\emph{M}_{bol}}$=0.125 for C-rich stars 
of mass $\sim$1.55~$\rm{M_{\odot}}$.  If we adopt the distant modulus to the 
LMC of 18.54 \citep{2006ApJ...642..834K}, the transition luminosity from the 
O-rich to C-rich stars is \[M_{\rm{bol}}=-3.57-0.6m_1~,\] based on the data 
of \citet{1990ApJ...352...96F} and \citet{1994ApJS...92..125V}. Matching the 
effective temperature for O-rich and C-rich stars at the transition
luminosity, the effective temperature of C-rich stars becomes
\begin{equation}
\label{cteff}
\rm{log~\emph{T}_{eff}=3.969+0.084\emph{m}_{1}+0.125\emph{M}_{bol}}~~.
\end{equation}

With $L=4\pi \sigma R^{2}T_{\rm{eff}}^{4}$ and
$\rm{log~{\emph{L}/L_{\odot}}=-0.4{(\emph{M}_{bol}-4.75)}}$,
we can derive the bolometric magnitude for O-rich and C-rich stars as a 
function of stellar radius $R$ from Equations (\ref{T-LM}) and (\ref{cteff}).
For O-rich stars,
\begin{equation}
\label{olum}
M_{\rm{bol}}=4.362-0.222m_{1}-3.704\rm{log}\emph{R}~~,
\end{equation}
and for C-rich stars,
\begin{equation}
\label{clum}
M_{\rm{bol}}=1.189-0.373m_{1}-2.222\rm{log}\emph{R}~~.
\end{equation}
If we substitute $R$ with $R_{\rm{min}}$ 
or $R_{\rm{max}}$, then we get the minimum or maximum bolometric magnitude 
of the red giants with just-detectable light variations or with Roche lobes 
that are just full, respectively.  We note that the formulae given
above for $T_{\rm eff}$ are derived from single star observations.  
However, these formulae should also give good results for red giants 
in binary systems as stellar radii are not greatly affected by the
presence of a binary companion \citep[][and 
references therein]{1971ARA&A...9..183P}.

\subsubsection{The MACHO and OGLE comparison data}\label{macho_ogle}
As noted in the introduction, \citet{1999IAUS..191..151W} and
\citet{2004AcA....54..347S} found that the fraction of red giants which
have detectable ellipsoidal variability is $\sim$0.5\% and 1--2\%,
respectively. The higher fraction found 
by \citet{2004AcA....54..347S} is due to the smaller photometric
errors of the OGLE observations and hence the ability of OGLE to
detect smaller ellipsoidal variations than MACHO.
The fraction of stars that show ellipsoidal variability
is an important input to our calculations, so we recompute these
fractions accurately. We also investigate the amplitude required for
detectability of ellipsoidal variability in each of these two studies.

Using the MACHO data of \citet{1999IAUS..191..151W} and selecting the top
one magnitude of the RGB, we find that the fraction of red giants with
detectable ellipsoidal variability is 0.61\%.  Similarly, using the 
OGLE\,II ellipsoidal variables on the top one magnitude of the RGB from
\citet{2004AcA....54..347S} and the corresponding total red giant
population from \citet{2000AcA....50..307U}, we find an ellipsoidal
fraction of 1.33\%.

\begin{figure}
\begin{center}
\includegraphics[angle=0,width=0.45\textwidth]{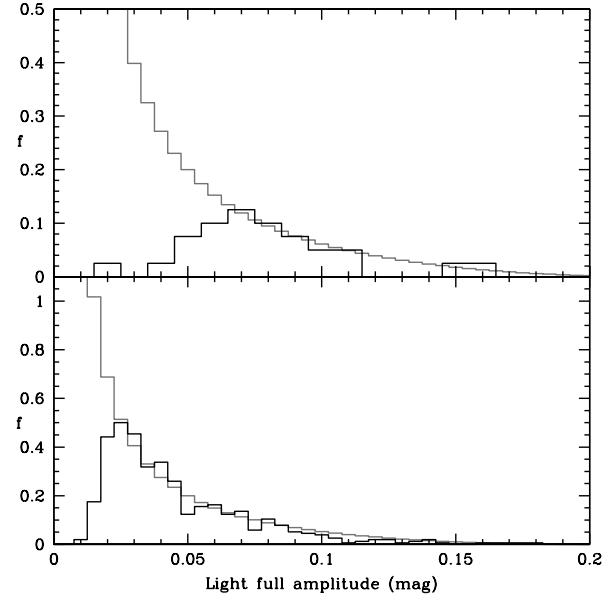}
\caption{The relative distribution of full light amplitude for ellipsoidal 
variables on the top one magnitude of the RGB.  Grey lines are estimates 
from the Monte Carlo simulation and the black lines are the observed OGLE\,II 
distribution from \citet{2004AcA....54..347S} (bottom panel) and the observed
MACHO distribution from \citet{1999IAUS..191..151W} (top panel). The histograms 
in the top panel apply for the MACHO red band while the histograms in the 
bottom panel apply for the OGLE\,II $I$ band.\label{fig_det_lim}}
\end{center}
\end{figure}

The distributions of full light amplitude for the MACHO and OGLE\,II
ellipsoidal variables are shown in Fig. \ref{fig_det_lim}.  It is clear 
that the OGLE\,II observations detect variability at considerably lower
amplitude than the MACHO observations.  Also shown in Fig. \ref{fig_det_lim}
are the corresponding distributions of ellipsoidal light amplitudes
on the top one magnitude of the RGB computed from our Monte Carlo model.
We take the detectability limits for ellipsoidal variability to be
the light amplitude at which the observed number of stars has fallen
to 50\% of the estimated true number. These limits are 0.055 magnitudes 
for the $M_R$ band data and 0.019 magnitudes for the OGLE $I$ band data.
We use these detectability limits and the observed relative fractions
of ellipsoidal variables to calibrate our Monte Carlo model.

\subsubsection{A summary of evolutionary outcomes}\label{outcome}
We aim to predict the fractions of binary and single red giant stars that 
suffer each of the various possible evolutionary fates. The possible 
evolutionary outcomes are itemized below.
\begin{enumerate}
\renewcommand{\theenumi}{(\arabic{enumi})}
\item The star reaches the AGB tip without filling its Roche lobe. Single 
stars will follow this path and produce a single PN. Wide and intermediate
period binaries will also follow this path, producing a PN with its nucleus 
in a wide or intermediate period binary.
\item  The star fills its Roche lobe on the AGB but above the RGB tip.  
There are three possible outcomes.
\begin{enumerate}
\item  A CE event leads to the formation of a close binary consisting
of an AGB star core and a secondary star.  In this case, evolution of 
the AGB star core leads to creation of a PN with a close binary nucleus.
\item  A CE event leads to a merger of the two stars with the secondary 
star merging into the red giant envelope.  The merged star will then 
evolve to the AGB tip as a single star.
\item Stable mass transfer occurs leading to the formation of
an intermediate period binary.  If all the mass lost from the red
giant was transferred to the secondary, no PN would be visible.
However, there is likely to be some mass lost from the system and
we assume a PN will be seen in the rare cases that stable mass transfer
occurs.
\end{enumerate}
\item  The star fills its Roche lobe below the RGB tip. 
For stars with initial masses
$m<1.85~\rm{M_{\odot}}$, the Roche lobe filling will occur on the RGB. 
However, for stars with $m>1.85~\rm{M_{\odot}}$, this will occur on the 
EAGB. There are three possible outcomes of such CE events.
\begin{enumerate}
\item  
A CE event leads to the formation of a close binary consisting
of a red giant core and a secondary star.  In this case, evolution of 
the red giant core to high effective temperatures is too slow to lead 
to creation of a PN (see the post-RGB evolution tracks of \citet*
{1998A&A...339..123D} but note the comments at the beginning of 
Section~\ref{stand_mod}; the cores of EAGB stars also evolve slowly as
they are burning He in a shell). We will generally call these stars 
post-RGB stars even though some of them come from the EAGB. Theoretical 
considerations suggest that the outcome of a CE event in a close binary 
on the RGB (or EAGB) is likely to be a post-CE binary consisting of a 
low-mass He-core (or CO-core) white dwarf and a basically unaltered 
low-mass secondary \citep{1976IAUS...73...75P,1984ApJ...277..355W}.
We suggest that some Population II Cepheids, UU Herculis variables and 
RV Tauri stars with luminosities below the RGB tip are also formed in 
this way. Most, if not all, RV Tauri stars are binary systems that have 
mid-infrared dust emission and circumbinary disks 
\citep{2005A&A...435..161D, 2007IAUS..240..682L}, as expected from 
terminating red giant evolution on the RGB or AGB by Roche lobe filling. 
Observationally, some of the Population II Cepheids, UU Herculis variables 
and RV Tauri stars in the LMC have luminosities below the RGB tip luminosity 
\citep{1998AJ....115.1921A}.  
\item  A CE event leads to a merger of the two stars with the secondary 
star merging into the red giant envelope. The merged star will then evolve 
to the AGB tip as a single star.
\item Stable mass transfer occurs leading to the formation of
an intermediate period binary.  Some of the binary Population II Cepheids, 
UU Herculis variables and RV Tauri stars with dust disks could be 
produced in this way.
\end{enumerate}
\end{enumerate}

\subsubsection[]{Simulation normalization}\label{norm_test}

We use the well determined fractions of the sequence E stars on the
top one magnitude of the RGB (Section \ref{macho_ogle}, Wood's 0.61\%
and Soszy\'nski's 1.33\%) to normalize our calculations. Given
our adopted IMF, period distribution and mass ratio distributions,
some binary stars will have parameters such that the Roche lobe fills
or nearly fills at these luminosities ($-2.6 > M_{\rm bol} > -3.6$),
and the star will exhibit detectable ellipsoidal light variations
characteristic of sequence E stars over a luminosity sub-interval of 
one magnitude.  The lifetime as a sequence E star, corresponding to
this luminosity sub-interval, is calculated by using the evolutionary
rate described in Section \ref {mass_loss}. Some of the other binaries
will have their evolution terminated by interaction before reaching
$M_{\rm bol} = -2.6$ and some will evolve through the full interval
without showing any detectable light variations.  In the latter case,
the lifetime is also calculated using the evolutionary rate described
in Section \ref {mass_loss}. We calculate and sum the sequence E
lifetimes of all RGB and EAGB stars in the interval $-2.6 > M_{\rm
  bol} > -3.6$ and we also calculate and sum the total lifetimes of all red
giants as they pass through this interval. The ratio of the summed
sequence E lifetimes of those binaries showing detectable ellipsoidal
variations to the total summed lifetimes of all binaries gives the apparent
fraction of sequence E stars. This apparent fraction due to binaries
alone is always greater than the observed fraction.  Therefore, after
we have considered $10^6$ binary stars in the Monte Carlo simulation,
we add single stars to the population of stars, assuming they evolve
through the full interval $-2.6 > M_{\rm bol} > -3.6$, until the
predicted and observed sequence E fractions match.  Note that we are
implicitly assuming here that the ratio of binary to single stars in
the LMC is a free parameter and that it can differ from the ratio estimated
for stars in the solar vicinity.  According to \citet{1991A&A...248..485D}
the percentage of solar-type stars that are single is $\sim$33\% 
while \citet{2010ApJS..190....1R}, who derive a period distribution
quite similar to that of \citet{1991A&A...248..485D}, estimate 
the percentage of single solar-type stars to be $\sim$56\%.

We emphasize that our main purpose in this study is to use the
observed sequence E star population, which has initial orbital periods
of $\sim$100-600 days, to estimate the population of red giants whose
evolution is affected by binary interactions.  These interacting stars
typically have initial orbital periods of $\la$3000 days.  The
observed period distribution of binary stars
\citep{1991A&A...248..485D, 2010ApJS..190....1R} is very broad with a
peak at $\sim$$10^5$ days and the stars we are primarily concerned
with occupy only a small part of the overall period distribution.  Our
predictions related to red giant evolution should thus be relatively
insensitive to the overall period distribution (provided it is
smooth).  However, estimates of the numbers of stars with periods
$>>$3000 days, and the number of single stars, will depend more
sensitively on the overall period distribution.

\section{Results for the standard model}\label{stand_mod}

In this section, we present the results for our standard model
using both the Wood and Soszy\'nski frequencies for sequence E stars.
The models produce the relative birthrates of stars that terminate
their AGB or RGB evolution. The numbers  in Table~\ref{results_tab}
give these relative birthrates.

Throughout this paper we make some assumptions. Firstly, we assume that
all stars that leave the AGB produce 
a PN.  Secondly, when we compare ratios
of populations of different types of PNe (for example close binaries PNe
and all PNe) we assume that the PN lifetimes is independent of the
AGB termination process (CE event or wind) although this may not be the case
\citep[e.g.][]{2006ApJ...650..916M} .  Similar considerations
apply for the post-RGB stars.  It should be kept in mind that we are really
comparing birth rates.  

An example of how evolution rates could be different for post-CE stars
and single or non-interacting binaries is provided by the class of
post-AGB (perhaps post-RGB) star that consists of a binary system with
a dusty circumbinary disk, an orbital period of a few hundred days or
more, and a large eccentricity \citep{2007BaltA..16..112V}.  These
stars could result from AGB or RGB termination by stable mass
transfer.  Alternatively, their evolutionary path could be caused by the
high eccentricity of their orbits.
The disks around these stars may provide a reservoir of hydrogen that
can be accreted back onto the high luminosity star to fuel nuclear burning for
an extended period, at the same time keeping the hydrogen envelope thick enough to
maintain a relatively low $T_{\rm eff}$  which is insufficient to excite a
PN.

At the other extreme is the possibility that a common envelope
event in a RGB star can remove enough of the hydrogen above the burning
shell that the star comes out of the CE event with $T_{\rm eff}$
large enough to excite a PN almost immediately \citep{2006ApJ...650..916M}.
In our modelling, we assume that this does not happen and that the
evolution of post-RGB stars is so slow that the circumstellar shell has dispersed
before $T_{\rm eff}$ reaches values high enough to produce a PN
(see Section~\ref{outcome}, item 3a).

\subsection{Fractions of binary planetary nebula nuclei and post-RGB stars}\label{fraction}
With the standard inputs and the evolutionary scenarios described above, we 
examine which binary systems form close binary PNe, intermediate period binary PNe,
wide binary PNe, single 
binary PNe, and post-RGB binaries. The predicted PNe fractions of the standard 
model are listed in the first two lines of Table \ref{results_tab}. In the total population of
planetary nebula nuclei (PNNe), close binaries make up 9\% or 7\%, intermediate period 
binaries make up 27\% or 23\%, wide binaries make up 55\% or 46\%, and 
single stars make up 3\% or 19\%, using Wood's (0.61\%) or Soszy\'nski's (1.33\%) frequency, respectively.  
The ratio of post-RGB star births 
to close binary PNNe births is $\sim$50\%. Of 
those binaries undergoing a CE event on the AGB, we find $\sim$1.4\% suffer a merger while 
the remainder produce a close binary. Due to the relatively higher binding energy of stars on the RGB,
the frequency of mergers happening on the RGB
is much higher than that on the AGB.  Below $M_{\rm bol} \sim -1.5$, essentially all
red giants with a main sequence companion merge when Roche lobe overflow occurs.
However, if the companion is a white dwarf, a double degenerate binary can be
produced down to low luminosities.
Our standard model predicts that merged stars,
which are likely to be rapidly rotating red giants, would make up 
about 5\% of the red giants on the RGB above $M_{\rm bol} \sim -1.5$. Lower on the giant branch, 
the merged fraction will be less than 5\%. 
\citet{2011ApJ...732...39C} estimate the observed fraction of rapidly 
rotating red giants to be $\sim$2\%. 

Our results can be compared to the binary evolution models of
\citet{1995MNRAS.272..800H}.  Their models 4 and 11 have input
parameters similar to our standard model except that our initial
period distribution is different from theirs (they assume a constant
number of binaries per interval of $\log a$, where $a$ is the binary
separation).  In our standard model, we find that about 8\% of PNNe
are close binaries whereas \citet{1995MNRAS.272..800H} predict that
about 4--5\% of PNNe are close binaries (see the column ``CE Ejection
(AGB)'' in Table 1 of \citet{1995MNRAS.272..800H} and column ``Binary
PNNe / CE'' in this paper).  The reason for our higher fraction can be
traced to the different initial period distributions.  Binary systems
that undergo CE ejections on the AGB have initial orbital periods of
about 230--1400 days or separations of about 230--760 R$_{\odot}$
(assuming the binary component masses are $\sim$1.5 M$_{\odot}$).  All
binary systems with initial periods (or separations) greater than
these values will produce PNe through wind mass loss, as will single
stars.  The number of binaries with initial periods (or separations)
that lead to a CE event on the AGB, relative to the number of
all binaries plus single stars that produce PNe, is roughly twice as large
in our standard model as in models 4 and 11 of
\citet{1995MNRAS.272..800H}.  This explains the higher fraction of
close binary PNNe produced in our standard model.  (We note that our
models 15 and 16, described in Section~\ref{non-standard}, have similar period
distributions to models 4 and 11 of \citet{1995MNRAS.272..800H}.  Both
sets of models predict that about 4--5\% of PNNe are close binaries
indicating similar outcomes for similar input physics.)

Finally, we note that the fractions of close binary PNNe given by
\citet{1995MNRAS.272..800H} are somewhat arbitrary since these
fraction depends on the maximum binary separation allowed for their
binary systems (they adopt, without giving justification, a maximum
separation of $5.75\times 10^6$ R$_{\odot}$).  They also make the assumption
that there are no single stars.  In our models, the fraction of stars
that are single or in wide or intermediate period binaries is
observationally constrained by the fraction of stars showing
ellipsoidal variability on the RGB.  

We now examine observational estimates of the fraction of close binary PNe. 
Searches carried out by Bond and collaborators for close binary central stars 
of PNe using photometric variability techniques obtained a fraction of 10--15\% for 
close binary PNe 
\citep*{1987fbs..conf..221B, 1992IAUS..151..517B,1994ASPC...56..179B,2000ASPC..199..115B}.
However, the survey bias is not well understood since the PNe samples were 
monitored over 30 years with many different observing campaigns 
\citep*{2008AJ....136..323D}. A new survey for close binary central stars 
by \citet{2009A&A...496..813M} got a fraction of 12--21\% for close binary PNe. 
This survey is more efficient and was carried out in a relatively uniform manner, 
by searching for periodic photometric variability of homogeneous PNe samples. 
However, their 12--21\% is not a definitive fraction for close binary central 
stars. For example, the four close binary central stars PHR 1744-3355, PHR 
1801-2718, PHR 1804-2645 and PHR 1804-2913 in the \citet{2009A&A...496..813M} 
sample of 22 have been subsequently questioned and excluded by 
\citet{2011apn5.confE.328M} which would reduce their close binary fraction to
10--16\%.  On the other hand, accounting for the effect of orbital inclination 
on the light curve amplitude, the derived close binary fraction could increase.

Our standard model prediction of $\sim$7--9\% for close binary PNe, is
lower than the fractions estimated from the observations mentioned
above. However, our simulation is strongly constrained by the observed
fractions of sequence E stars. In addition, our model reproduces the
observed light amplitude, period and velocity amplitude
distributions of sequence E stars as well as the fraction of low
mass He white dwarfs in the total white dwarf population (see
Section \ref{wd_mdist}).  This suggests that our estimated fraction
of PNe with close binary central stars is reliable.

\begin{landscape}
\begin{table}
 \caption{Model parameters and relative birthrates of red giant progeny\label{results_tab}}
 \begin{tabular}{lccccccccccccccccccccc}
 \hline
 \multicolumn{9}{c}{\bf{Model parameters}} & & \multicolumn{8}{c}{\bf{PNNe (\%)}}& &\multicolumn{2}{c}{\bf{Post-RGB stars (\%)}} &  \\
 \cline{1-9} \cline{11-18} \cline{20-21} 
 & & & & & & & & & & \multicolumn{4}{c}{\bf{Single PNNe}}& &\multicolumn{3}{c}{\bf{Binary PNNe}}& &\multicolumn{2}{c}{\bf{Binaries}} &\\
 \cline{11-14} \cline{16-18} \cline{20-21}
 No& $\alpha_{\rm CE}$& $P_{\rm max}$& $R_{\rm burst}$& IMF& P$_{\rm dist}$& $q_{\rm dist}$& B & $q_{\rm crit}$& &
 Single & Wide & MgAGB & MgRGB && IntP & Stable & CE && CE & Stable & PRGB/PN \\
 \hline
 \multicolumn{22}{l}{\bf{Standard model}} \\
 1 & 1.0 & 500 &   10 & S1955  & DM1991             & DM1991     &   0 &   H57 &w&  3.31 & 54.90 &  0.12 &  5.71 && 27.20 &  0.01 &  8.74 &&100.00 &  0.00 & 0.0448 \\
   &     &     &      &        &                    &            &     &       &s& 18.52 & 46.24 &   0.11&  4.83 && 22.97 &  0.01 &  7.31 &&100.00 &  0.00 & 0.0375 \\
 \multicolumn{22}{l}{\bf{One parameter variation models}}\\
 2 & 0.6 & 500 &   10 & S1955  & DM1991             & DM1991     &   0 &   H57 &w&  1.67 & 55.00 &  0.55 &  7.28 && 27.21 &  0.01 &  8.28 &&100.00 &  0.00 & 0.0275 \\
   &     &     &      &        &                    &            &     &       &s& 17.18 & 46.28 &  0.46 &  6.10 && 22.95 &  0.01 &  7.01 &&100.00 &  0.00 & 0.0233 \\
 3 & 0.3 & 500 &   10 & S1955  & DM1991             & DM1991     &   0 &   H57 &w&  0.00 & 54.99 &   1.83&  8.81 && 27.31 &  0.01 &  7.05 &&100.00 &  0.00 & 0.0101 \\
   &     &     &      &        &                    &            &     &       &s& 16.62 & 45.97 &   1.50&  7.29 && 22.73 &  0.01 &  5.88 &&100.00 &  0.00 & 0.0085 \\\
 4 & 1.0 &$+\infty$&10& S1955  & DM1991             & DM1991     &   0 &   H57 &w& 19.77 &  0.00 &  0.13 &  5.89 && 65.38 &  0.01 &  8.81 &&100.00 &  0.00 & 0.0461 \\
   &     &     &      &        &                    &            &     &       &s& 33.00 &  0.00 &   0.11&  4.86 && 54.59 &  0.01 &  7.44 &&100.00 &  0.00 & 0.0388 \\
 5 & 1.0 & 100 &   10 & S1955  & DM1991             & DM1991     &   0 &   H57 &w&  0.00 & 68.71 &   0.13&  5.80 && 16.62 &  0.01 &  8.72 &&100.00 &  0.00 & 0.0441 \\
   &     &     &      &        &                    &            &     &       &s& 14.55 & 59.29 &   0.11&  4.80 && 13.93 &  0.01 &  7.31 &&100.00 &  0.00 & 0.0370 \\
 6 & 1.0 & 500 &    1 & S1955  & DM1991             & DM1991     &   0 &   H57 &w&  3.08 & 55.46 &   0.07&  5.71 && 27.99 &  0.08 &  7.61 &&100.00 &  0.00 & 0.0650 \\
   &     &     &      &        &                    &            &     &       &s& 18.67 & 46.68 &   0.06&  4.78 && 23.35 &  0.07 &  6.41 &&100.00 &  0.00 & 0.0546 \\
 7 & 1.0 & 500 &   10 & Reid02 & DM1991             & DM1991     &   0 &   H57 &w&  1.10 & 56.35 &   0.19&  5.59 && 27.63 &  0.01 &  9.12 &&100.00 &  0.00 & 0.0407 \\
   &     &     &      &        &                    &            &     &       &s& 15.48 & 48.04 &   0.17&  4.75 && 23.72 &  0.01 &  7.83 &&100.00 &  0.00 & 0.0342 \\
 8 & 1.0 & 500 &   10 & S1955  & Flat               & DM1991     &   0 &   H57 &w& 62.15 &  3.91 &  0.11&  10.45 && 16.37 &  0.01 &  6.99 &&100.00 &  0.00 & 0.0480 \\
   &     &     &      &        &                    &            &     &       &s& 66.87 &  3.42 &   0.11&  9.10 && 14.39 &  0.01 &  6.11 &&100.00 &  0.00 & 0.0419 \\
 9 & 1.0 & 500 &   10 & S1955  & DM1991             &$n(q)=q^{-0.4}$&0 &   H57 &w& 11.40 & 41.73 &   0.20&  6.61 && 30.57 &  0.02 &  9.47 &&100.00 &  0.00 & 0.0533 \\
   &     &     &      &        &                    &            &     &       &s& 23.55 & 36.11 &   0.17&  5.68 && 26.32 &  0.02 &  8.15 &&100.00 &  0.00 & 0.0461 \\
10 & 1.0 & 500 &   10 & S1955  & DM1991             & $n(q)=1$   &   0 &   H57 &w& 27.55 & 34.43 &   0.09&  4.93 && 25.37 &  0.03 &  7.61 &&100.00 &  0.00 & 0.0447 \\
   &     &     &      &        &                    &            &     &       &s& 35.73 & 30.60 &   0.08&  4.37 && 22.50 &  0.02 &  6.70 &&100.00 &  0.00 & 0.0398 \\
11 & 1.0 & 500 &   10 & S1955  & DM1991             & $n(q)=q$   &   0 &   H57 &w& 37.00 & 30.58 &   0.03&  3.69 && 22.62 &  0.04 &  6.05 &&100.00 &  0.00 & 0.0439 \\
   &     &     &      &        &                    &            &     &       &s& 41.04 & 28.62 &   0.02&  3.43 && 21.13 &  0.04 &  5.71 &&100.00 &  0.00 & 0.0411 \\
12 & 1.0 & 500 &   10 & S1955  & DM1991             & DM1991     & 500 &   H57 &w&  8.16 & 52.63 &   0.10&  5.33 && 27.32 &  0.73 &  5.74 && 99.30 &  0.70 & 0.0478 \\
   &     &     &      &        &                    &            &     &       &s& 19.65 & 45.97 &   0.09&  4.71 && 23.96 &  0.63 &  4.98 && 99.28 &  0.72 & 0.0420 \\
13 & 1.0 & 500 &   10 & S1955  & DM1991             & DM1991     &   0 &   0.8 &w& 23.07 & 43.35 &   0.10&  4.42 && 21.79 &  0.72 &  6.55 && 92.18 &  7.82 & 0.0367 \\
   &     &     &      &        &                    &            &     &       &s& 26.68 & 41.35 &   0.10&  4.26 && 20.74 &  0.68 &  6.19 && 92.02 &  7.98 & 0.0351 \\
 \multicolumn{22}{l}{\bf{Multiple parameter variation models}} \\
14 & 1.0 & 500 &   10 & S1955  & R2010              & $n(q)=1$   &   0 &   H57 &w& 16.38 & 42.63 &   0.10&  4.77 && 28.20 &  0.03 &  7.88 &&100.00 &  0.00 & 0.0450 \\
   &     &     &      &        &                    &            &     &       &s& 26.15 & 37.63 &   0.08&  4.23 && 24.92 &  0.02 &  6.96 &&100.00 &  0.00 & 0.0399 \\
15 & 1.0 & 100 &    1 & MS1979 & Flat               & $n(q)=1$   &   0 &   H57 &w& 63.43 &  8.96 &  0.05&  11.00 && 10.79 &  0.10 &  5.67 &&100.00 &  0.00 & 0.0720 \\
   &     &     &      &        &                    &            &     &       &s& 66.63 &  8.13 &  0.05&  10.12 &&  9.73 &  0.12 &  5.23 &&100.00 &  0.00 & 0.0650 \\
16 & 1.0 & 100 &    1 & MS1979 & Flat               & $n(q)=q$   &   0 &   H57 &w& 71.36 &  7.44 &   0.01&  7.96 &&  8.91 &  0.13 &  4.19 &&100.00 &  0.00 & 0.0654 \\
   &     &     &      &        &                    &            &     &       &s& 71.99 &  7.29 &   0.02&  7.71 &&  8.66 &  0.13 &  4.21 &&100.00 &  0.00 & 0.0630 \\
17 & 1.0 & 500 &   10 & S1955  & DM1991             & DM1991     & 500 &   0.8 &w& 26.15 & 42.00 &   0.08&  4.24 && 22.16 &  1.59 &  3.77 && 86.64 & 13.36 & 0.0420 \\
   &     &     &      &        &                    &            &     &       &s& 28.44 & 40.62 &   0.08&  4.17 && 21.44 &  1.56 &  3.68 && 86.06 & 13.94 & 0.0396 \\
\hline
\end{tabular}
{\bf Model parameters:}
$\alpha_{\rm CE}$ is the orbital energy transfer efficiency (see Section~\ref{out_CE});
$P_{\rm max}$ is the maximum initial period (in years) for which 
a binary is assumed to influence the shape of PNe; $R_{\rm burst}$
is the star formation rate burst factor (see Section~\ref{generation}); IMF is the assumed initial mass function 
(`S1955' from \citealt{1955ApJ...121..161S}, `Reid02' from \citealt{2002AJ....124.2721R}, 
`MS1979' from \citealt{1979ApJS...41..513M}); P$_{\rm dist}$ is the initial period 
distribution (`DM1991' from \citealt{1991A&A...248..485D}, `R2010' from 
\citealt{2010ApJS..190....1R}, `Flat' means $dN \propto d{\rm log}P$); 
$q_{\rm dist}$ is the initial mass ratio distribution $n(q) \propto dN/dq$; 
B is the parameter in the tidally enhanced mass loss rate formula (see 
Section.~\ref{mass_loss}); $q_{\rm crit}$ = ($m_2/m_1$)$_{\rm crit}$ is the critical
mass ratio above which Roche lobe overflow is stable (`H57' is the value given by
equation (57) of \citealt{2002MNRAS.329..897H}).\\
{\bf Model results:}
Lines containing `w'/`s' use Wood's/Soszy\'nski's sequence E frequency to normalize the results. 
The numbers in columns 11--19
give the relative birth rates for PNNe and post-RGB stars (as a percentage).  The sum of all
PNNe birthrates is 100\%, as is the sum of the post-RGB star birthrates.  
The ratio of the total post-RGB star birthrate
to the total of the PNNe birth rate is given column 20.
PNNe types are: stars that are born single (Single);
wide binaries which have $P > P_{\rm max}$ and which do no influence the shape of their PNe 
(Wide); binary stars that merge on the AGB (MgAGB) or the RGB (MgRGB); 
intermediate period binaries that never fill their Roche lobes but which have $P < P_{\rm max}$ 
(IntP); binary stars that fill their Roche lobes
and then undergo stable mass transfer (Stable); and binary stars that fill their Roche lobes
and then undergo a CE event leaving a close binary (CE).
Post-RGB are: those that result from CE events (CE); those that result
from stable mass transfer (Stable).
\end{table} 
\end{landscape}

\subsection{Properties of sequence E stars}\label{seqE}

To further test our simulation, we predict the binary properties of
sequence E stars and compare them with observations where possible.
The main available data are the photometric variations from MACHO and
OGLE experiments, which provide a probe into the period distribution.
\citet{2010MNRAS.405.1770N} provide radial velocity observations of
sequence E stars, although only 11 binary systems were studied.

\subsubsection{Period distribution}

\begin{figure}
\begin{center}
\includegraphics[angle=0,width=0.40\textwidth,height=0.850\hsize]{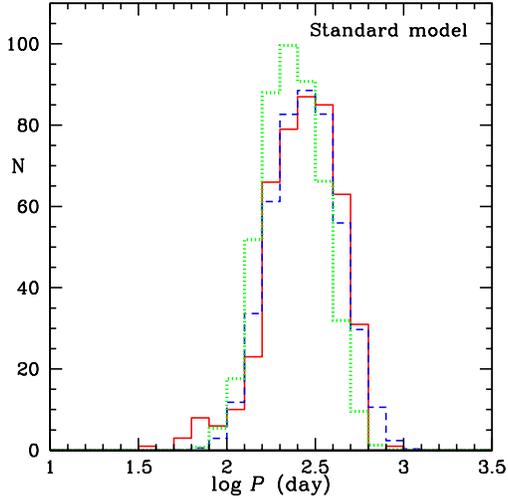}
\caption{The orbital period distribution of sequence E stars. The red solid 
line denotes the distribution from the observations of \citet{2004AcA....54..347S},  
the green dotted line denotes the distribution from the model using Wood's 
frequency, and the blue short dashed line denotes the distribution from the 
model using Soszy\'nski's frequency.\label{fig_Eperiod}}
\end{center}
\end{figure}

In Fig. \ref{fig_Eperiod}, we present the orbital period distribution
of sequence E stars, predicted with Wood's and Soszy\'nski's
frequencies.  The orbital period contribution of a single binary to
the distribution is the average of the periods at onset and end of
detectable sequence E variability on the RGB, weighted by the sequence
E lifetime for the system: note that the orbital period typically
changes by less than 3\% during the sequence E phase (due to mass loss
and tidal interaction).  We also show the distribution of the
observed periods, using the sample of \citet{2004AcA....54..347S}. We
only consider the sequence E stars on the top one magnitude of the
RGB, so the stars in the observed period distribution are selected
based on their luminosity. Fig. \ref{fig_Eperiod} shows that the
simulation reproduces the observation very well: not only are the
total numbers of sequence E stars reproduced, as required, but their period
distribution is reproduced as well. We also
note that the shape of the light amplitude distribution is reproduced
well by the model (Fig.~\ref{fig_det_lim}).

\subsubsection{Full velocity amplitude distribution}
\begin{figure}
\begin{center}
\includegraphics[angle=0,width=0.42\textwidth,height=0.85\hsize]{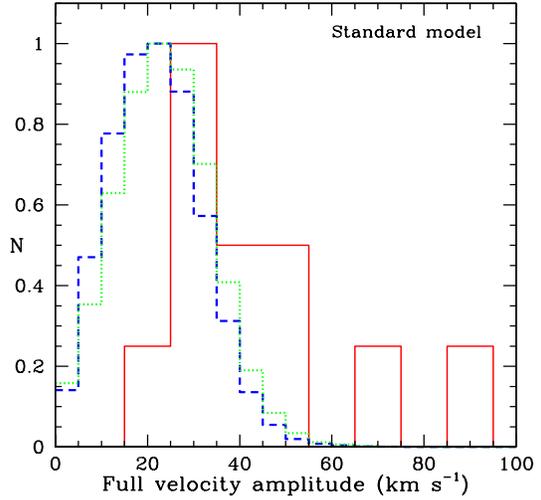}
\caption{The full velocity amplitude distribution of sequence E stars. The red 
solid line denotes the distribution from the observation of \citet{2010MNRAS.405.1770N}
with 11 stars, the green dotted line denotes the distribution from the model using Wood's 
frequency, and the blue short dashed line denotes the distribution from the 
model using Soszy\'nski's frequency. The distribution peaks are normalized to 1. \label{fig_Evelocity}}
\end{center}
\end{figure}

In Fig. \ref{fig_Evelocity}, we show the full velocity amplitude distributions 
from our model and the radial velocity observations by \citet{2010MNRAS.405.1770N}. 
The expected full velocity amplitude concentrates in the range of $\sim$5 to 50 
km s$^{-1}$, with a peak at $\sim$22.5 km s$^{-1}$. It shows a reasonable agreement with 
the observations, although the observed number is small.

\subsubsection{Mass distribution}
\begin{figure}
\begin{center}
\includegraphics[angle=0,width=0.45\textwidth,height=0.85\hsize]{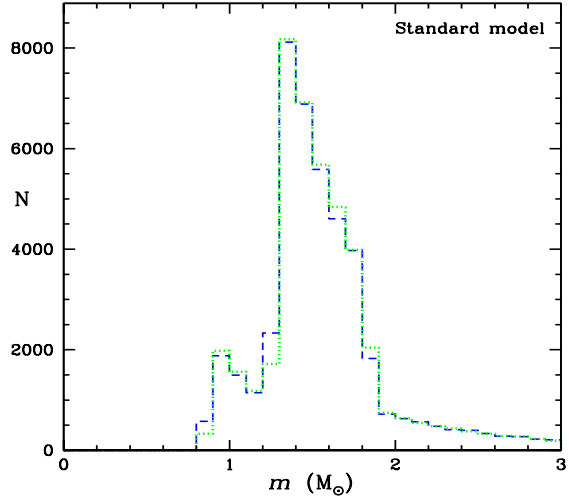}
\caption{The mass distribution of the sequence E stars. The green dotted line 
denotes the distribution from the model when using Wood's frequency while the 
blue short dashed line denotes the distribution from the model when using 
Soszy\'nski's frequency.\label{fig_Emass}}
\end{center}
\end{figure}

In Fig. \ref{fig_Emass}, we show the mass distribution of the sequence E stars. 
The mass range is from 0.85--3.0~$\rm{M_{\odot}}$, with the LMC star burst 
beginning at 1.3~$\rm{M_{\odot}}$. The lower limiting mass of 0.85~$\rm{M_{\odot}}$ 
is set by the mass of the oldest stars and stellar wind mass loss. There are a 
large number of stars in the range of 1.3--1.85~$\rm{M_{\odot}}$, where the latter 
mass is the maximum mass for stars that ascend the RGB with an electron degenerate 
core. The more massive stars in Fig. \ref{fig_Emass} are EAGB stars at luminosities
corresponding to the top magnitude of the RGB.

\subsubsection{Mass ratio distribution}
\begin{figure}
\begin{center}
\includegraphics[angle=0,width=0.45\textwidth,height=0.85\hsize]{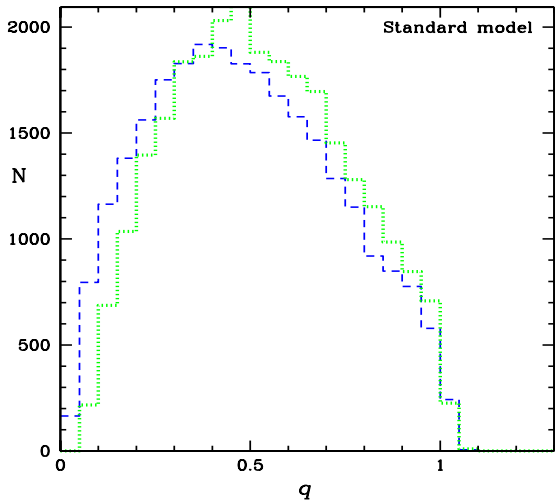}
\caption{The same as Fig. \ref{fig_Emass} but for the mass ratio.
\label{fig_Eratio} }
\end{center}
\end{figure}

In Fig. \ref{fig_Eratio}, we show the mass ratio distribution of sequence E 
stars. The mass ratio is less than 1.1, with a peak at $\sim$0.5. Mass loss 
from the red giant is the reason that some systems have mass ratio slightly 
greater than 1.

\subsubsection{Orbital separation distribution}
\begin{figure}
\begin{center}
\includegraphics[angle=0,width=0.45\textwidth,height=0.85\hsize]{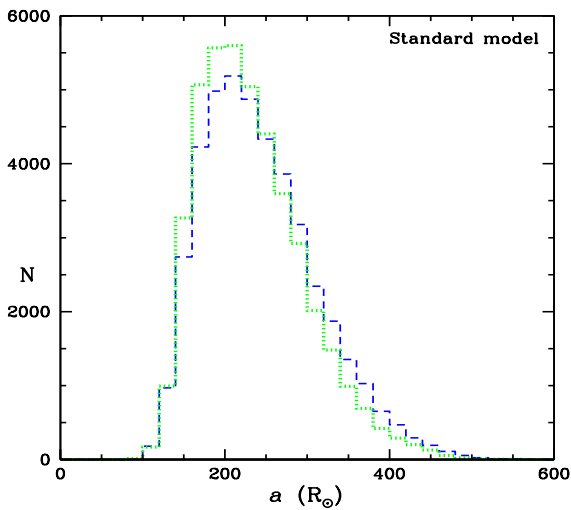}
\caption{The same as Fig. \ref{fig_Emass} but for the orbital separation.
\label{fig_Eseparation} }
\end{center}
\end{figure}

In Fig. \ref{fig_Eseparation}, we show the orbital separation distribution of sequence E 
stars. The separation is in the range of $\sim$100--600 $\rm{R_{\odot}}$, with 
a narrow peak at $\sim$200~$\rm{R_{\odot}}$. 

\subsubsection{Ellipsoidal variability lifetime}

\begin{figure}
\begin{center}
\includegraphics[angle=0,width=0.45\textwidth,height=0.85\hsize]{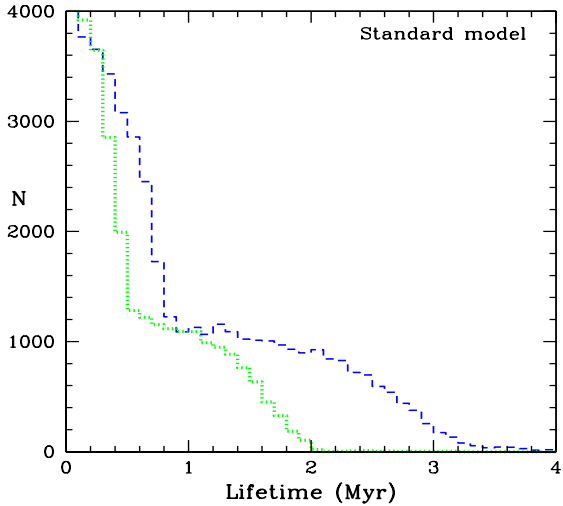}
\caption{The same as Fig. \ref{fig_Emass} but for the lifetime of the ellipsoidal 
variability.\label{fig_Elifetime}}
\end{center}
\end{figure}

We show the predicted ellipsoidal variability lifetime of sequence E
stars in Fig. \ref{fig_Elifetime}. The distribution starts at a peak
of $\sim$0.2 Myr due to the contribution of EAGB stars with
$m>1.85~\rm{M_{\odot}}$, and then falls to zero at 2 or 4 Myr. 
Longer lifetimes are predicted in the Soszy\'nski case than the Wood
case.  The reason is that OGLE II has a more sensitive detection limit
for photometric variations ($\sim$0.02 mag) than MACHO ($\sim$0.05 mag) (see
Fig.~\ref{fig_det_lim}), so Soszy\'nski's stars are detected initially
with less-filled Roche lobes and it then takes longer before the lobe
is filled and RGB evolution is terminated. The average lifetime for
sequence E stars is $\sim$0.95 Myr, when the ellipsoidal variation
amplitude is more than 0.02 magnitudes.

\citet{2007BaltA..16....1M,2011arXiv1110.2361M} claims that in
symbiotic stars, the red giant radius is often close to half the Roche
lobe radius yet ellipsoidal light variations are seen in these stars
with an amplitude that suggests the Roche lobe is almost filled.  If
this claim is correct, then it suggests some unidentified source of
light variation that simulates an ellipsoidal variation.  The sequence E stars are not
as extreme as the symbiotic stars where the red giant is usually a
semi-regular or Mira variable with a large mass loss rate and the
companion is a hot accreting white dwarf whose radiation may affect
the facing surface of the red giant.  We therefore do not expect our
ellipsoidal light curve calculations for sequence E stars to be
affected by an unidentified source of ellipsoidal-like light
variation.  Nevertheless we investigate how such an extra source would
alter our results if it did exist.  To simulate the extra light
amplitude suggested in symbiotic stars, we replaced the Roche lobe
filling factor $f$ in Equation~(\ref{ramp}) by $2f$.  The effect of this
is to make detectable ellipsoidal variability occur when the radius of
the red giant is half the usual value.  Note that this does not affect
the evolution of any star, it simply means that ellipsoidal
variability is detectable in red giants with wider orbits than in our
usual calculations.  This in turn causes a larger fraction of binaries
to appear as sequence E stars.  Now, in order to reproduce the
observed fractions of sequence E stars, a much larger population of
single stars is required than in our usual models.  For example, in
our standard model, about 10\% of stars are single and 90\% are in
binaries whereas in the models with the modified Equation~(\ref{ramp})
about 80\% of stars are single and 20\% are in binaries.  This means
that the fraction of all stars that undergo CE events or mergers is
reduced by a factor of $90/20 = 4.5$.  In particular, the fraction of
close binary PNNe is reduced from $\sim$8\% to $\sim$1.8\%.  (We also
note that in our modified models the ellipsoidal light amplitudes of
stars nearly filling their Roche lobes are increased by a factor of 16
giving rise to ellipsoidal light curve amplitudes up to 3 magnitudes,
which are never observed (see Fig.~\ref{fig_det_lim}).  Of course, the
modified Equation~(\ref{ramp}) has no physical basis so we cannot really
say what the ellipsoidal amplitudes should be.)  

\subsection{Properties of binary PNNe and post-RGB stars}\label{CEsystem}

\subsubsection{Luminosity distribution}
\begin{figure}
\begin{center}
\includegraphics[angle=0,width=0.45\textwidth,height=1.4\hsize]{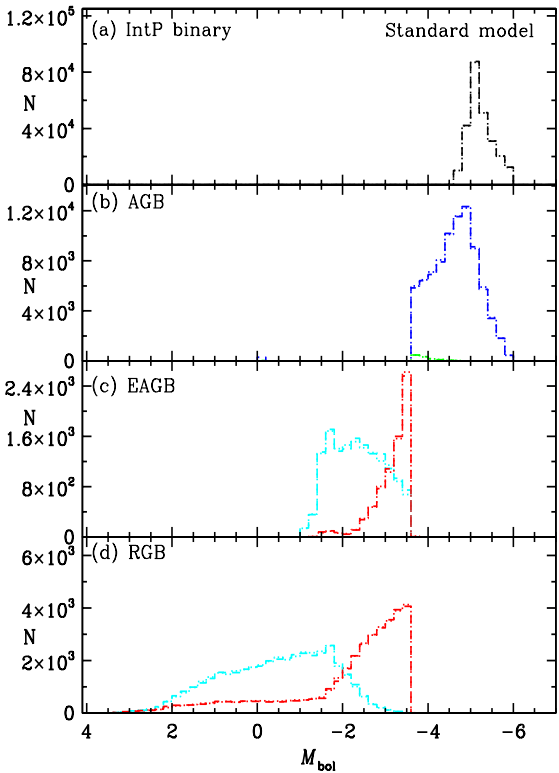}
\caption{Luminosity distributions of binary red giants at the termination of their 
red giant evolution. Panel (a): intermediate period 
binary PNNe with $P<500$ years. Panel (b): 
close binary PNNe (blue) and merged AGB stars (green). Panel (c): EAGB binaries 
(red) and merged EAGB stars (cyan/grey). Panel (d): RGB binaries (red) and 
merged RGB stars (cyan/grey). In all panels, dotted lines denote the distributions 
from the model when using Wood's frequency, short dashed lines denote the 
distributions from the model when using  Soszy\'nski's frequency. 
\label{fig_lum}}
\end{center}
\end{figure}

In Fig. \ref{fig_lum}, we show the luminosity distributions of the outcomes of 
the binary red giants according to the standard model. As expected, intermediate period binary 
PNNe have luminosities of the AGB tip, close binary PNNe have AGB luminosities 
above the RGB tip, and EAGB and RGB binaries undergoing a CE event but not merging have 
luminosities below the RGB tip. Mergers occur preferentially at 
luminosities lower than the luminosities of CE events that lead to binaries. 
The RGB binaries formed at $M_{\rm bol} \ga -1.5$ are double degenerates.
In these systems, the small radius of the companion white dwarf allows them
to avoid a merger.

\subsubsection{Mass distribution}\label{wd_mdist}
\begin{figure}
\begin{center}
\includegraphics[angle=0,width=0.45\textwidth,height=1.4\hsize]{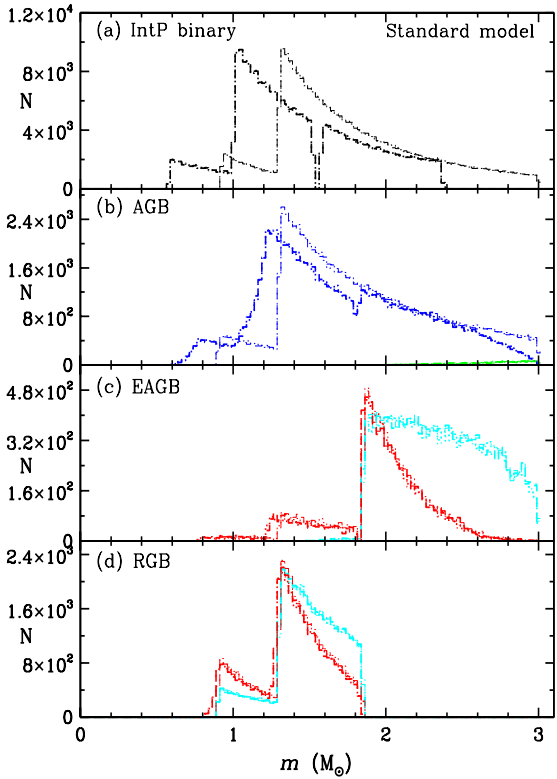}
\caption{Mass distributions of the binary red giants according to their evolutionary
fates. Thin lines are for the initial mass distributions. Thick lines are for the mass 
distribution at the termination luminosity: the top panel shows the mass for
intermediate period binaries at the AGB tip and just before the beginning of the 
superwind phase; the other panels show the mass at the start of the CE event. 
Colors are as in Fig. \ref{fig_lum}.
\label{fig_mass1}}
\end{center}
\end{figure}

In Fig. \ref{fig_mass1}, we show the mass distributions of the binary red giants
at their termination luminosities. As expected, stars in intermediate period binaries 
lose a significant amount of their mass 
via the Reimers-like stellar wind. Mergers occur preferentially at lower luminosity 
so such systems have the least mass loss.

\begin{figure}
\begin{center}
\includegraphics[angle=0,width=0.45\textwidth,height=0.87\hsize]{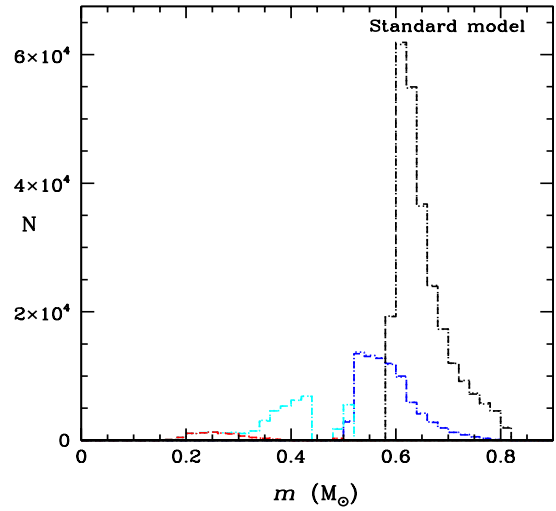}
\caption{Mass distributions of intermediate period binary PNNe (black), 
close binary PNNe (blue),
post-RGB and post-EAGB binaries (cyan/grey) and double degenerate secondaries (red).
Note that the cyan curves include the double degenerate component.
\label{fig_mass2}}
\end{center}
\end{figure}

At the end of the superwind phase or a CE event, a red giant has lost its whole envelope, leaving a He 
(or CO) core as a remnant. In Fig.~\ref{fig_mass2}, we present the mass distribution 
after the red giant envelope has been lost. The mass of the close binary PNNe is in the 
range of $\sim$0.50--0.80 $\rm{M_{\odot}}$,  with most of the stars concentrated 
between 0.50--0.65 $\rm{M_{\odot}}$. For the post-RGB binaries, the mass is  
lower, with a range of $\sim$0.2--0.45~$\rm{M_{\odot}}$. The  post-EAGB CO-cores 
concentrate around 0.50~$\rm{M_{\odot}}$. The mass distributions of 
intermediate period binary PNNe 
are also shown in Fig.~\ref{fig_mass2}. The masses are in the range of 
$\sim$0.55--0.85~$\rm{M_{\odot}}$, with a strong peak at $\sim$0.6$~\rm{M_{\odot}}$.

Mass determination of 86 PNNe by \citet{1990A&A...240..467S} shows a mass distribution 
similar to our prediction: the mass range is 0.55--0.75 $\rm{M_{\odot}}$, and most of 
the PNNe concentrate between 0.55--0.65 $\rm{M_{\odot}}$. Similarly, 
\citet{1993ApJS...88..137Z} determined the mass of 303 PNNe, reporting a mass distribution 
from 0.55--0.85 $\rm{M_{\odot}}$, with a narrow peak at 0.6 $\rm{M_{\odot}}$. The population 
synthesis model results of \citet{1993ApJ...418..794Y} and \citet{1995MNRAS.272..800H} give 
mass distributions of close binary PNNe (with $\alpha_{\rm CE} = 1$) that are similar to ours. 

For the post-RGB stars, the predicted masses are consistent with
observations of low-mass white dwarf binaries. For example,
\citet*{2005ApJS..156...47L} and \citet{2011MNRAS.413.1121R} determined
the mass distribution of local white dwarfs and found a distribution
with a peak at 0.4 $\rm{M_{\odot}}$ as well as a more dominant peak
around 0.6 $\rm{M_{\odot}}$ which corresponds to CO white dwarfs
produced by evolution off the AGB. The low
mass peak is produced predominantly by CE events on the RGB,
although some of these stars ($<$30\%) may be single 
\citep[][and references therein]{2010ApJ...708..411K,2011ApJ...730...67B,2011MNRAS.413.1121R}
which requires an exotic type of evolution such as an unusually high
wind mass loss on the RGB, or ejection of the envelope by a giant
planet, or mergers of two He white dwarfs (we do not find any such
events).  The studies of \citet{2005ApJS..156...47L} and 
\citet{2011MNRAS.413.1121R} suggest that the birth rate of low
mass He white dwarfs is about 4-10\% of the total white dwarf birth
rate.  Our standard model predicts this relative birth rate to be
9.6\%, in good agreement with the observations.  Our predicted mass
distribution is also consistent with the observations and the simulations of
\citet{1993ApJ...418..794Y} and \citet{1995MNRAS.272..800H}.  
Note that the He white dwarfs with masses from 0.2--0.3
$\rm{M_{\odot}}$ in Fig.~\ref{fig_mass2} are nearly all double
degenerates and they form about 19\% of the He white dwarfs.  
Some He white dwarfs are known to be in double degenerate systems
\citep*{1995MNRAS.275..828M} although the observed fraction is very 
poorly known \citep{2007ApJ...671..761K}. Our models predict that
only $\sim$0.1\% of sequence E stars will have a white dwarf companion
(on the way to producing a double degenerate), consistent with our finding
mentioned in the introduction that there is no evidence for degenerate
companions in a sample of 110 sequence E stars.

Finally, we note that our adopted star formation history means that
there are no stars with initial masses greater than 3.0\,M$_{\odot}$.
Even if we were to allow recent star formation by (say) using $f(t) =
b$ for $t < 0.5$, the relative number of stars with $M >
3.0$\,M$_{\odot}$ would be tiny because of the combined effects of the
IMF and the star formation history.  We would not simply be extending
the plots in Figs.~\ref{fig_mass1} to higher masses according to the
IMF.  We would be extending the plots at a level reduced by a factor of
10 in height since the stars with mass 1.3--3.0\, M$_{\odot}$ were
produced during the burst.

\subsubsection{Period distribution}
\begin{figure}
\begin{center}
\includegraphics[angle=0,width=0.45\textwidth,height=0.9\hsize]{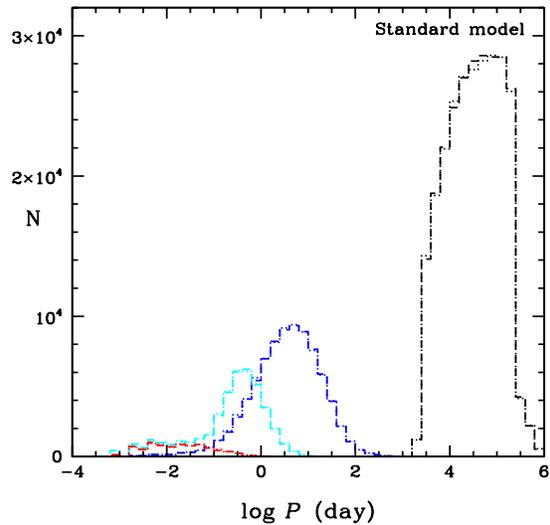}
\caption{Orbital period distributions of the objects shown in Fig.~\ref{fig_mass2}.
Line types are the same as in that figure.
\label{fig_period}}
\end{center}
\end{figure}

In Fig. \ref{fig_period}, we present the orbital period distribution of the close binary
PNNe and post-RGB binaries. The orbital periods of the close binary PNNe have a range 
from 0.01--1000 days, and a peak at $\sim$4 days. The orbital periods of post-RGB 
binaries are between 0.01--10 days, and have a narrow peak at $\sim$0.3 days. 

Orbital period distributions of close binary PNNe predicted by
\citet{1993ApJ...418..794Y} and \citet{1995MNRAS.272..800H} have a
similar distribution to ours, showing a period range of few hours to
1000 days. Observationally, periods of close binary PNNe are all less than
16 days
\citep{2000ASPC..199..115B,2009PASP..121..316D,2009A&A...496..813M}
with the current period distributions peaking at or below $P \sim 1$
day.  This peak is at shorter periods than predicted by the model (the blue
curve in Fig.~\ref{fig_period}).  The absence of the long period PNNe
in the current sample could be due to the observational bias against
long period stars: in variability searches such as that of
\citet{2009A&A...496..813M}, short periods in close binaries are
needed to get the detectable light amplitudes from ellipsoidal and
reflection effects; and in radial velocity searches, close binaries
with short periods are needed to get the high velocity variations
needed for these hot compact stars which have relatively broad lines.
Alternatively, the current observed period distribution and the 
model period distributions can be brought into better agreement
by using a lower value of $\alpha_{\rm CE}$ (see 
Section~\ref{non-standard}).

The orbital period distribution of local binary white dwarfs
\citep[e.g.][]{2008MNRAS.390.1635R,2010A&A...520A..86Z} also peaks
at periods less than or near 1 day.  The model prediction
given by the cyan line in Fig.~\ref{fig_period} has a peak around 1 day
but more stars with periods longer than 1 day than the current
observational estimates.

\subsubsection{Mass ratio distribution}
\begin{figure}
\begin{center}
\includegraphics[angle=0,width=0.45\textwidth,height=0.9\hsize]{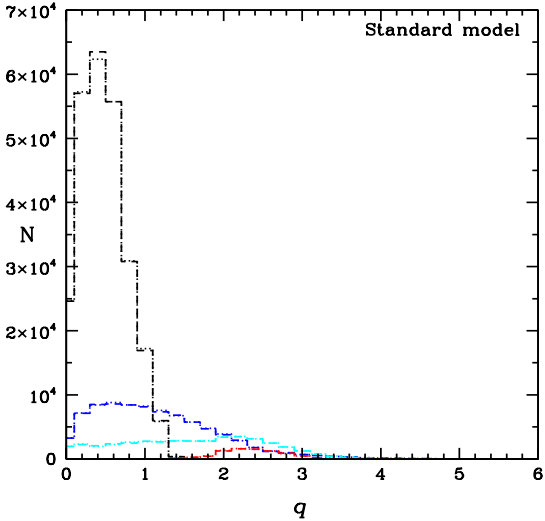}
\caption{Mass ratio distributions of the objects shown in Fig.~\ref{fig_mass2}.
Line types are the same as in that figure.
\label{fig_ratio}}
\end{center}
\end{figure}

In Fig. \ref{fig_ratio} we show the mass ratio distributions of close PNNe and 
post-RGB binaries. This is similar to the prediction of \citet{1995MNRAS.272..800H}.

\subsection{The population of PNe in the LMC}\label{pn_pop}

We use our model to predict the number of PNe in the
inner 25 square degrees of the LMC searched for PNe by
\citet{2006MNRAS.373..521R}.  As part of our modelling procedure to
estimate the apparent fraction of sequence E stars in the top one
magnitude of the RGB ($-2.6 > M_{\rm bol} > -3.6$), we compute the
average lifetime $\tau_{\rm 1mag}$ in this magnitude interval for all
RGB and EAGB stars which evolve into this interval.  We find
$\tau_{\rm 1mag} = 2.77 \times 10^6$ years for the standard model,
with a deviation from this value of less than 3\% for other models.  
The average lifetime is determined by the evolution rate and variation
in the average lifetime is caused by the small fraction of stars 
whose lives are terminated by CE events in this interval.
From the SAGE catalog \citep{2006AJ....132.2268M}, we obtained the near
and mid-infrared photometric observations of all stars in the area 
searched by \citet{2006MNRAS.373..521R}.  Using the position of the giant branch
in the (J,J$-$[3.6]) diagram, we selected all the stars (RGB and EAGB)
in a parallelogram corresponding to the top one magnitude of the RGB.
The parallelogram has sides J=13.25, J=14.25, J$-$[3.6]=$3.828-0.17{\rm J}$ 
and J$-$[3.6]=$3.328-0.17{\rm J}$ (see \citet{2006AJ....132.2034B} for 
SAGE colour-magnitude diagrams for the whole LMC).
We found the total number of stars $n_{\rm RGB+EAGB} = 58356$ and
estimate an error of less than 5\% in this number.  The uncertainty is
due to scattering of stars into and out of the parallelogram 
in the (J,J$-$[3.6]) diagram because of observational errors and 
contamination by stars at the red end of core helium burning loops. 
Confusion is not a problem for these relatively bright stars 
\citep[see][]{2000ApJ...542..804N,2006AJ....132.2034B}.

The final quantity we need in order to estimate the number of 
PNe in the inner 25 square degrees of the LMC is the
average lifetime $\tau_{\rm PN}$ of LMC PNe.  We note that
there is a steep drop in the number of LMC PNe with diameters 
between 7 and 8 arcsec \citep[see Fig.~16 of][]{2006MNRAS.373..521R} 
so we adopt 7.5 arcsec as a typical maximum diameter
for LMC PNe.  If we use a distance modulus to the LMC of 18.54 
\citep{2006ApJ...642..834K} and a PN expansion velocity $v_{\rm exp}$ 
of 34.5 km s$^{-1}$ \citep[the median for the 94 LMC PNe 
in][]{1988ApJ...327..639D} then we find $\tau_{\rm PN} = 2.6 
\times 10^4$ years.  Using the above numbers, we 
predict that the number of PNe in the region surveyed by
\citet{2006MNRAS.373..521R} will be $n_{\rm RGB+EAGB}\times
\tau_{\rm PN} / \tau_{\rm 1mag} = 548$ PNe. 
The estimated number of PNe is in good agreement with
the 541$\pm$89 PNe observed by \citet{2006MNRAS.373..521R}.  Since
more than 90\% of these PNe come from single or non-interacting binary
stars in our model, this means that most such stars produce a PN.
This is contrary to the ``Binary Hypothesis''
\citep{2006ApJ...650..916M,2009PASP..121..316D} which suggests that
binary interaction is required to produce a PN.

There are a small number of AGB star terminations that may not produce
a detectable PN.  Most of the single or non-interacting binary stars
in our model terminate their AGB evolution by a superwind.  However, some
have their evolution terminated by the low mass loss rate Reimers wind.  These
are the stars with the lowest initial mass $0.9 < {\rm M}/M_{\odot} < 0.92$ which make up
only 0.7\% of the AGB terminations.  It is these stars that may not produce
a detectable PN \citep{2005AJ....130.2717S}.

\section{Results for models with varied input assumptions}\label{non-standard}

\begin{figure}
\begin{center}
\includegraphics[angle=0,width=0.45\textwidth,height=0.85\hsize]{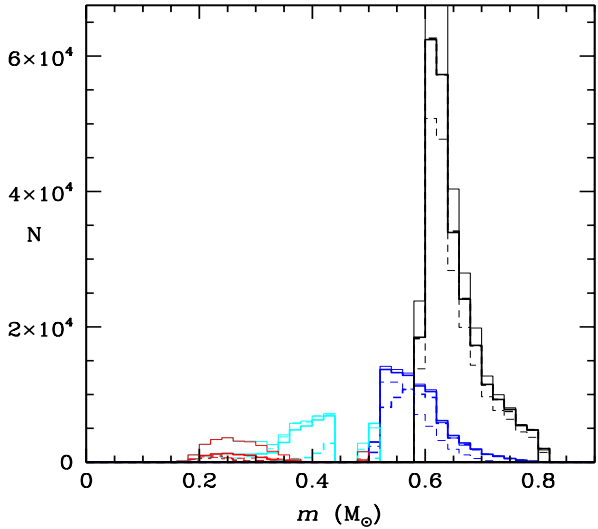}
\caption{Mass distributions of intermediate period binary PNNe (black), 
close binary PNNe (blue),
post-RGB and post-EAGB binaries (cyan/grey) and double degenerate secondaries (red) in
models 1 (thick solid lines), 3 (thick dashed lines), 11 (thin solid lines) and 12 (thin dashed lines).
Results are shown only for the models using Wood's sequence E frequency.
\label{multifig_mass2}}
\end{center}
\end{figure}

\begin{figure}
\begin{center}
\includegraphics[angle=0,width=0.45\textwidth,height=0.9\hsize]{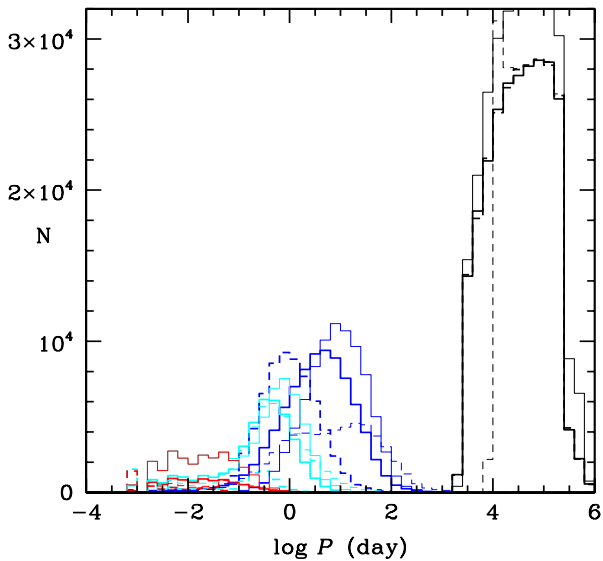}
\caption{Orbital period distributions of the objects shown in Fig.~\ref{multifig_mass2}.
Line types are the same as in that figure.
\label{multifig_period}}
\end{center}
\end{figure}

In order to investigate how dependent the results are on the model 
parameters (e.g. $\alpha_{\rm CE}$, IMF, period and mass ratio distribution), 
as well as to compare the results with other population synthesis calculations, we 
have run models with adjusted 
input parameters.  Models have been made varying one, two or three of the parameters
described above and the results are shown in Table~\ref{results_tab}.
We discuss the effects on each model of varying
the model parameters.  For selected models, distributions of final masses
and periods are plotted in Figs.~\ref{multifig_mass2} and \ref{multifig_period},
which are similar to Figs.~\ref{fig_mass2} and \ref{fig_period}, respectively.

\vspace{3mm}\noindent {\it Models 2 and 3: the energy efficiency parameter $\alpha_{\rm CE}$.}\\
The value of the energy efficiency parameter $\alpha_{\rm CE}$ is
controversial and there is no consensus for its value. As noted in
Section~\ref{out_CE}, a value of 1.0 seems consistent with some
observational constraints.  However, smaller values of $\alpha_{\rm
  CE}$ have also been derived along with estimates of dependency on the
parameters of the binary
\citep{2007ApJ...665..663P,2010A&A...520A..86Z,
  2011MNRAS.411.2277D,2011ASPC..447..115D}.  To investigate the effect 
of changes to $\alpha_{\rm CE}$ on our modelling results, the standard value
of 1.0 was changed to 0.6 and 0.3 in models 2 and 3, respectively.  As
expected, decreasing $\alpha_{\rm CE}$ causes more mergers to occur at
the expense of fewer CE events (Table~\ref{results_tab}).  Reducing
$\alpha_{\rm CE}$ from 1 to 0.3 changes the fraction of close binary
PN produced from $\sim8$\% to $\sim6.5$\%.  The results for model 3
are shown in Figs.~\ref{multifig_mass2} and \ref{multifig_period}.
The most interesting thing about these plots is that the periods
of the close binary PNNe now peak just shortward of 1 day, in much
better agreement with the current observed period distribution.
A lower value of $\alpha_{\rm CE}$ is a way of removing the discrepancy
between the observed period distribution and the period distribution
predicted by our standard model.

\vspace{3mm}\noindent {\it Models 4 and 5: The upper limiting period $P_{\rm max}$.}\\ 
$P_{\rm max}$ has been changed from the standard value of 500 years to
infinity or 100 yrs.  The latter value was used by
\citet{1995MNRAS.272..800H}. The main effect of this parameter is to
shift stars between the intermediate period, wide and single
categories.

\vspace{3mm}\noindent {\it Model 6: The star burst ratio $R_{\rm burst}$.}\\
In this model, the star burst ratio $R_{\rm burst}$ in Section
\ref{generation} was changed to 1.0 to simulate a constant star
formation rate, which is often used in models
\citep{1990ApJ...358..189D,1992A&A...261..188D,1993ApJ...418..794Y,
  1995MNRAS.272..800H,2007ApJ...665..663P}.  The change produces more
stars of mass less than 1.3\,M$_{\odot}$ relative to higher mass stars.
The effect is a small reduction in the number of CE events on the AGB
and an increase in the number of CE events on the RGB.  This is
probably due to the smaller initial orbital separations of the lower
mass binaries, given that the same period distribution is assumed at
all masses.

\vspace{3mm}\noindent {\it Model 7:The IMF.}\\
Here the Salpeter IMF was changed to the \citet*{2002AJ....124.2721R} 
power law (with an exponent of $-1.3$).  The effect on the birthrate ratios is
small, with slightly more CE events on the AGB and fewer on the RGB.

\vspace{3mm}\noindent {\it Model 8: The period distribution $P_{\rm dist}$.}\\
Here a flat period distribution with $dN\propto{d\,\rm{log}\emph{P}}$
was used.  This period distribution has been commonly used 
\citep[see][]{1990ApJ...358..189D,1992A&A...261..188D,1995MNRAS.272..800H}.
The period distribution has a large influence on the
relative birthrates of single, wide and intermediate period PNNe.
Since the flat period distribution has far fewer stars born at long
periods relative to the periods of the sequence E stars, it is
necessary to add many more single stars than in the standard model in
order the get the fraction of sequence E stars on the top magnitude of
the RGB to agree with the observational value.  The fraction of stars
undergoing CE events is slightly lower than in the standard case.
This model is a good indicator of the fact that using sequence E
stars to normalize our results keeps the fraction of CE events in
our simulations fairly constant when model parameters are changed.

\vspace{3mm}\noindent {\it Models 9 to 11: Mass ratio distributions $q_{\rm dist}$.}\\
Initial mass ratio distributions $n(q)=Cq^a$ with $a=-0.4$, 0 or 1
were tried.  The first exponent was that found by
\citet{2007A&A...474...77K} for main-sequence A and B stars ($M \sim
2-15$\,M$_{\odot}$) in the Sco OB2 association.  The higher the
fraction of high mass companions, the fewer the number of mergers, as
might be expected from the higher orbital energy available to eject
the red giant envelope.  The higher the fraction of high mass companions
also gives a smaller percentage of CE events in our model.  This is
probably because, for a given fraction of sequence E stars, a higher
mass companion can give a detectable ellipsoidal light variation at a
larger orbital separation, reducing the likelihood of a CE event
occurring.  The final mass and period distributions for model 11 are shown
in Figs.~\ref{multifig_mass2} and \ref{multifig_period}.  These
plots clearly show an increase in the fraction of double degenerates
produced when the number of high mass companions is increased.  The
periods of the close binary PNNe are also increased slightly.

\vspace{3mm}\noindent {\it Model 12: Tidally enhanced mass loss.}\\
The tidally enhanced mass loss rate of \citet{1988MNRAS.231..823T} was
tried, with the free parameter `$B$' set to 500, although its actual
value is very uncertain
\citep{1995MNRAS.277.1443H,1998MNRAS.296.1019H,1998ApJ...496..842S,
  2000MNRAS.316..689K,2001A&A...367..513F,2002MNRAS.329..897H}.
The main effect is an increase in the number of stars undergoing stable
mass transfer (because of the increase in $m_2/m_1$ resulting from mass
loss from the red giant) and a decrease in the number of CE events.
The final mass and period distributions for model 12 are shown
in Figs.~\ref{multifig_mass2} and \ref{multifig_period}.  The reduced
number of CE events on the AGB is clearly evident in the figures.
There is also an extension of the period distribution of close binary
PNNe to higher values, further increasing the discrepancy between
the observed and predicted period distributions.  The enhanced mass 
loss also significantly increased the minimum period of intermediate period PNNe.

\vspace{3mm}\noindent {\it Models 13 and 17: Critical mass ratio $q_{\rm crit}$.}\\
The critical mass ratio $q_{\rm crit} = (m_2/m_1)_{\rm crit}$, above which mass
transfer from a Roche lobe filling red giant is stable, has some uncertainty.
\citet{2008MNRAS.387.1416C} investigated the value of $q_{\rm crit}$ appropriate
for mass transfer from red giants under various conditions.  For the more luminous stars
and for mass transfer efficiencies more than 0.5, the value
of $q_{\rm crit}$ is mostly similar to that used here in our standard
model.  However, values as low as $(m_2/m_1)_{\rm crit} = 0.8$ are found by
\citet{2008MNRAS.387.1416C} in some cases.  In order to see how dependent our models are on
$q_{\rm crit}$, we ran a simulation with $q_{\rm crit} = 0.8$.  We also
ran a simulation with $q_{\rm crit} = 0.8$ and `$B$' = 500 to see how the extra
reduction in the red giant mass from the enhanced mass loss would affect the number of stars 
undergoing stable mass transfer.  As expected,
the number of stars undergoing stable mass transfer is greatly increased,
especially on the RGB.  The fraction of binaries undergoing CE events is 
decreased because of the alternative evolutionary path for Roche lobe filling
stars.  Changing $q_{\rm crit}$ to 0.8 makes no significant difference to the
mass and period distributions.

\vspace{3mm}\noindent {\it Model 14: The \citet{2010ApJS..190....1R} distributions.}\\
Here we use the period and mass ratio distributions of \citet{2010ApJS..190....1R} which can
be considered as updated versions of the distributions of \citet{1991A&A...248..485D}.
\citet{2010ApJS..190....1R} have a period distribution similar to that of
\citet{1991A&A...248..485D} but they have a flat mass ratio distribution.  The results
are very similar to those of model 10 which also has a flat mass ratio distribution.

\vspace{3mm}\noindent {\it Models 15 and 16: The \citet{1979ApJS...41..513M} IMF.}\\
Here we use the \citet{1979ApJS...41..513M} lognormal IMF appropriate for low-mass stars,
combined with a flat period distribution, and $n(q) = 1$ or $q$.  Our model 15 has similar
parameters to model 4 of \citet{1995MNRAS.272..800H}.
As with model 8, the flat period distribution leads to a high single star fraction in our simulation.
The combined effect of all parameters is to decrease the number of CE events on
the AGB and increase the number on the RGB.

\section{Conclusions}

We have used the population of red giants that are ellipsoidal
variables in the LMC to predict the fates of binary red giants. 
In our standard model, we found 7--9\% of PNe contain close
binaries, 23--27\% contain intermediate period binaries ($P < 500$
years) capable of influencing the shape of a PN, 46--55\% contain wide
binaries, 5--6\% are from merged stars and 3--19\% are from single
stars.  The production rate of post-RGB stars, consisting of a He
white dwarf and a low-mass secondary, is $\sim$50\% of the production
rate of close binary PNNe.

Our predicted fraction of close binary PNe (7-9\%) is somewhat lower
than current observational estimates of 10-20\%.  Similarly, the
observed orbital period distribution for close binary central stars of
PNe has few stars with periods more than 1 day whereas
the models predict that there should be a significant population of
binaries with periods out to 100 days.  The periods predicted by the
models can be brought into better agreement with the currently
observed periods by reducing $\alpha_{\rm CE}$.  The current
observational samples are small and detection techniques work best for
short period systems.  Larger samples and different detection
techniques are needed to improve the significance of the current
observational estimates.

The mass distribution of all white dwarfs produced shows two peaks
around 0.4 and 0.6 M$_{\odot}$ corresponding to He and CO white
dwarfs, respectively.  The birth rate of He white dwarfs is predicted
to be about 10\% of the total white dwarf birth rate, in agreement
with the observed birth rate ratio.

We estimate that about one third of PNe contain binaries capable of
influencing the shape of the nebula (the close and intermediate period
binaries), a significantly smaller fraction than the observed fraction
of non-spherical PNe which is $\sim$80\%.  This indicates that
mechanisms other than binarity are responsible for shaping PNe, such
as stellar rotation, magnetic fields or planets \citep[see][for a
  review]{2009PASP..121..316D}.

Using our model and the observed number of red giant stars on the top
one magnitude of the RGB, we predict that the number of PNe in the 
central 25 square degrees of the LMC should be 548.  This is
in good agreement with the 541$\pm$89 PNe observed by
\citet{2006MNRAS.373..521R}.  This result suggests that nearly all low
mass stars produce a PN in contrary to the ``Binary
Hypothesis'' \citep{2006ApJ...650..916M,2009PASP..121..316D} which
suggests that binary interaction is required to produce a PN.

We have also predicted the orbital element distributions of sequence E
stars. The predicted light amplitude, orbital period and full
velocity amplitude distributions agree well with the observations,
thus providing support for our modelling procedure.

\section*{Acknowledgements}
The authors would like to thank the referee, Maxwell Moe, for his careful
reading of this paper, leading to significant improvements.
The authors have been partially supported during this work by Australian Research 
Council Discovery Project DP1095368. JDN is also supported by the China Scholarship 
Council (CSC) student scholarship and the National Natural Science Foundation of 
China (NSFC) through grant 10973004.

\label{lastpage}

\end{document}